# Bibliometrics in Press.
# Representations and Uses of Bibliometric Indicators in the Italian Daily Newspapers.

*Eugenio Petrovich*
University of Siena

**Abstract:** Scholars in science and technology studies and bibliometricians are increasingly revealing the performative nature of bibliometric indicators. Far from being neutral technical measures, indicators such as the Impact Factor and the h-index are deeply transforming the social and epistemic structures of contemporary science. At the same time, scholars have highlighted how bibliometric indicators are endowed with social meanings that go beyond their purely technical definitions. These *social representations of bibliometric indicators* are constructed and negotiated between different groups of actors within several arenas. This study aims to investigate how bibliometric indicators are used in a context, which, so far, has not yet been covered by researchers, that of *daily newspapers*. By a content analysis of a corpus of 583 articles that appeared in four major Italian newspapers between 1990 and 2020, we chronicle the main functions that bibliometrics and bibliometric indicators played in the Italian press. Our material shows, among other things, that the public discourse developed in newspapers creates a favorable environment for bibliometrics-centered science policies, that bibliometric indicators contribute to the social construction of scientific facts in the press, especially in science news related to medicine, and that professional bibliometric expertise struggles to be represented in newspapers and hence reach the general public.

**Keywords:** bibliometrics ● science in the press ● Impact Factor ● h-index ● social construction of bibliometric indicators ● performance-based research evaluation ● Italy

# Introduction

With the advent of performance-based research evaluation systems in several countries all over the world, bibliometric indicators, such as the Journal Impact Factor and the h-index, have gained a prominent place in the "epistemic living space" of scientists (Biagioli and Lippman, 2020; Felt and Červinková, 2009; Hicks, 2012). The increasing role of metrics in academia can be understood as part of a broader set of organizational controlling techniques originating from the application of Neo-liberal New Public Management frameworks to higher education and R&D policies (Elzinga, 2012; Weingart, 2005). The "metricization of academia" (Burrows, 2012) reflects also the diffusion of the *audit culture* in contemporary societies, which stresses increasing demands of accountability in the national governance of the public sector (Dahler-Larsen, 2012; Strathern, 2000). As key components of these wider societal and research policy trends, bibliometric indicators have thus spread extensively among research administrators and funding agencies (Gläser and Laudel, 2007; Rijcke et al., 2016; Wouters, 2014), notwithstanding strong criticisms from the scientific community (e.g., (Alberts, 2013; Edwards and Roy, 2017; Seglen, 1997)) and manifestos such as the San Francisco Declaration on Research Assessment (DORA, 2013).

From a sociological point of view, one of the most interesting characteristics of bibliometric indicators is their capacity to *circulate in different contexts and communities of users* (Leydesdorff et al., 2016). Bibliometric indicators are not the prerogative of a single community, not even that of professional bibliometricians, but are shared between multiple groups of actors (Gläser and Laudel, 2007; Jappe et al., 2018).

Such a wide circulation can be appreciated by considering the various types of documents and inscriptions where bibliometric indicators appear. First, there is the scientific literature produced by researchers in the fields of bibliometrics, scientometrics, and library and information science, which is published in specialized journals such as *Scientometrics* or *JASIST*. In this context, the technical aspects of indicators are anatomized, new indicators developed, and recommendations for their correct use and interpretation issued (Kosten, 2016). However, since bibliometric indicators serve primarily to assess the impact in other science fields, also scientists from various disciplines contribute significantly to the discussion. For instance, biomedicine journals are an important arena for the discussion of the Impact Factor (e.g., (Misteli, 2013; Seglen, 1997)), whereas *Nature* published the Leiden Manifesto for the responsible use of metrics (Hicks et al., 2015). The circulation of bibliometric indicators, moreover, is not limited to scientific outlets. They appear in institutional reports prepared by governmental research evaluation agencies (e.g. (ANVUR, 2018)), in the documents made by commercial companies selling bibliometric data (e.g. (Elsevier, 2018)), and in the myriad of minor bureaucratic documents produced in the context of universities audit procedures (Hammarfelt and Rushforth, 2017).

In the light of their complex circulation among actors and texts, bibliometric indicators have been conceptualized as "boundary objects" (Star and Griesemer, 1989) whose meaning cannot be reduced to the technical definitions found in bibliometrics handbooks (Leydesdorff et al., 2016). The *social meaning* of the bibliometric indicators is instead constantly constructed and reframed in the processes of translation and negotiation that take place within and between the different communities that use and discuss them (Gläser and Laudel, 2007; Leydesdorff et al., 2016). Note that, consistently with the model of "boundary object" originally advanced by Star and Griesemer (Leigh Star, 2010), these groups do not need to share a consensual view on the validity of bibliometric indicators. Scientists, for instance, produce their own "folk theories" of bibliometric indicators, which overlap only partially with the theories advanced by bibliometricians (Aksnes and Rip, 2009). The lack of consensus, however, does not hinder the circulation of bibliometric indicators. In fact, it is one of its drivers.

## Communities involved in the social construction of bibliometric indicators

According to Leydesdorff, Wouters, and Bornmann (2016), the discourse about bibliometric indicators is mainly shaped by the translation processes among four groups:[1] a) the *producers* of bibliometric data and indicators. This group includes both private companies, such as Clarivate Analytics and Elsevier, that produce and maintain the citation indexes, and dedicated research centers, such as the Center for Science and Technology Studies (CWTS) in Leiden, that provide technically advanced bibliometric reports; b) the intellectual community of *bibliometricians*, who develop applied and basic research on indicators and other bibliometric topics and publish in specialized journals; c) the *research managers* who routinely order bibliometrics-based performance assessments of their institutions; d) the *scientists* being evaluated through bibliometrics. Usually, practicing scientists are not interested in bibliometrics *per se.* However, driven by the necessity to quantitatively demonstrate their research performance, many of them keep track of their bibliometric indicators on dedicated services such as Google Scholar or the software *Publish or Perish*.

---

[1] Leydesdorff and colleagues classification almost overlaps with that advanced by Gläser and Laudel (2007), who list: users in science policy and science management, the commercial owners of citation databases, the community of professional bibliometricians, and "amateur bibliometricians", identified as those academics, managers, and politicians who apply bibliometrics «without having the necessary professional background» (102). The notion of "amateur bibliometrics" was criticized by Hammarfelt and Rushfort (2017) because alleged amateur bibliometricians may show an awareness of the limitations of bibliometric methods comparable to that of their professional counterparts.

Specific tensions can be individuated between these communities (Leydesdorff et al., 2016). For instance, bibliometricians have repeatedly denounced the technical flaws of the h-index and advised against its use (see the next paragraph). Nonetheless, the h-index remains an indicator commonly employed by research managers and scientists because it is easy to compute and interpret compared to more sophisticated bibliometric indicators.

The study by Leydesdorff, Wouters, and Bornmann offers a useful starting point for understanding the different roles of the groups involved in the construction of the bibliometric indicators' social meaning. However, it can be improved in three respects to become a fully effective tool for describing empirical cases.

First, it does not consider that the *groups' weights* can be differently distributed in the national contexts. The situation in a country such as the Netherlands, where the bibliometric community is strongly institutionalized and professionalized (Petersohn and Heinze, 2018), cannot be equated with the situation in countries where the field of bibliometrics is poorly developed. Moreover, bibliometrics is not equally integrated into the research performance evaluation systems of the different countries around the world (Geuna and Martin, 2003; Hicks, 2010).

Secondly, *further communities* may participate in the public discourse on bibliometric indicators, in addition to those listed by Leydesdorff and colleagues. For instance, the journalists who write about university and research policy topics, as well as the scientists who intervene on these topics in the newspapers, play a crucial role in shaping the representation the public opinion has of bibliometric indicators. In this sense, public opinion and further stakeholders should be included in the model.

Thirdly, and closely related to the second point, the schema does not fully consider the *specificities of the various arenas* where the social construction of meaning and the negotiations between communities take place. A scientist writing an editorial on the Impact Factor for a prestigious medical journal is likely to adopt a different communicative strategy if she is interviewed on the same topic by a journalist for a daily newspaper.

In the light of these observations, this study aims to deepen our understanding of the social construction of bibliometric indicators' meaning and the communities involved therein by investigating an arena that have not yet been considered in the literature, namely the *arena of the generalist daily press*. We aim to investigate how bibliometric indicators are *represented* in newspapers, what *function* they play in press articles, and how journalists and other authors writing in newspapers *make sense* of them.

The rest of the paper is structured as follows. In the next section, the role of the press in the communication of science is reviewed and the rationale for focusing on newspapers' representation of bibliometrics is furtherly delineated. Then, the two indicators this study will focus on, the Impact Factor and the h-index, are briefly presented, and the research questions that will guide the analysis laid out. As the press is a very vast and differentiated arena, this study focuses on a specific context, namely the Italian press. In the following paragraph, the reasons for this choice are explained. In the Methodology section, the methods used for gathering the newspaper articles and analyze their content are described. Then, we present empirical results, which are organized into ten main themes. Lastly, in the Conclusions, we sum up the main findings.

## Bibliometrics in the press

The literature on the public communication of science convincingly shows that the newspapers, and the media more generally, perform «an extremely complex role in the communication of science, a role that in certain circumstances may act as a 'filter' and an 'arena' for researchers in specialist and inter-specialist areas» (Bucchi and Mazzolini, 2003: 53). Research has shown that the communication between science and the public cannot be understood as a linear process, where the public is the passive receiver of scientific knowledge coming from scientists through the popularization of science journalists (Bauer and Bucchi, 2007; Bucchi, 1998). In fact, the activity of communicating knowledge to external audiences constitutes a key element in the complex construction of scientific facts (Bucchi, 1998: 11). The public arena may even constitute the stage where scientific facts are first put into being and later dissolved, as the case of cold fusion shows (Bucchi, 1998: 3).

At present, no detailed study of the representation and function of bibliometric indicators in the press exists. This study aims at closing this gap. In particular, we are interested in detailing, among other things, how bibliometric expert knowledge, i.e., the knowledge developed by professional bibliometricians, is translated and actively elaborated in the medium of the daily newspapers. In this sense, we are not interested in pursuing a fact-checking or "debunking" approach (Bucchi, 1998). We do not aim to judge how "good" the journalists report the items of bibliometric knowledge they use. At any rate, it would be pointless to evaluate newspaper articles with the same standards used for specialist contributions. Rather, we are interested in describing how bibliometric knowledge is *translated and elaborated* in the press, also charting those cases where bibliometric concepts are stylized or simplified.

By investigating both the *representation* and the *use* of bibliometric indicators in the press, this study contributes to an emerging research stream in social studies of science that focuses on *bibliometric indicators in action*. So far, research has investigated bibliometric indicators in the contexts of applicants evaluation (Hammarfelt and Rushforth, 2017) and practices of knowledge production in the sciences (Castellani et al., 2016; Müller and de Rijcke, 2017; Rushforth and de Rijcke, 2015) and humanities (Hammarfelt and de Rijcke, 2015). We add to these studies the analysis of the *press context*.

## Research questions

This study analyzes the representation of bibliometrics in newspapers, with a special focus on the function and uses of the two most popular and discussed bibliometric indicators, the Journal Impact Factor (IF) and the h-index (Glänzel et al., 2019; Jappe et al., 2018; Todeschini and Baccini, 2016).

The IF was originally introduced in the 1960s by Eugene Garfield, a founding father of bibliometrics, and it is calculated annually by a commercial company, Clarivate Analytics (formerly Thomson Reuters). It corresponds to the ratio between a) the number of citations received in a given year by the documents published in a journal during the two previous years, and b) the total number of items published in that journal over the two previous years (Larivière and Sugimoto, 2019). Roughly, it can be understood as the average number of citations gathered by the articles published in a journal. The Impact Factor is commonly used as a measure of the prestige or quality of a scientific journal (Moed, 2005), despite the vast critical literature that has addressed its technical limitations and misuses (Vanclay, 2012). A recent survey of the perception researchers have of the IF shows that the prevailing attitude is ambivalent. In some fields, such as mathematics and statistics, researchers have mostly a negative opinion of it (Buela-Casal and Zych, 2012). The IF is in any case deeply entrenched in the epistemic culture of some areas, especially in the life sciences, where it can condition the choice of the research topics or collaboration strategy (Müller and de Rijcke, 2017; Rijcke et al., 2016).

The h-index, on the other hand, is a much younger indicator. As noted above, it was introduced by the physicist Jorge Hirsch (an outsider from the point of view of professional bibliometrics) to measure the scientific output of individual scientists. Differently from the IF, which is a journal metric, the h-index targets the individual researcher. A scientist has h-index $h$ if $h$ of her or his $P$ articles have at least $h$ citations and the remaining $P – h$ articles have received less than $h$ citations each (Hirsch, 2005). Since its introduction, a huge stream of research, both inside and outside bibliometrics, has focused on the h-index's properties (Schubert and Schubert, 2019). Professional bibliometricians have pointed out many of its technical flaws but these critical remarks did not stop the ascent and diffusion of the index, also in research evaluation systems.

By studying the representation of these two indicators in the press, as well as the image of bibliometrics more generally, we aim to answer questions among which are:

- How does the weight of different indicators in the press change over time? Do trends correlate with changes in research policies? Is news about research policy the main channel through which indicators reach the public?
- What functions do bibliometric indicators play in the rhetorical structure of newspaper articles?
- What are the most common attributes they are associated with? Can a "folk theory" of bibliometrics be delineated in the press? How are items of bibliometric knowledge (e.g., the definitions of IF and h-index or their technical limitations) translated and elaborated in the medium of the press?
- What are the entities the bibliometric indicators are mostly attributed to? Are they institutions or individual scientists? Are there scientific fields that are commonly mentioned in association with bibliometrics and bibliometric indicators?
- Who are the authors of the articles mentioning bibliometrics or bibliometric indicators? Are they mainly journalists or scientists participate in the discussion as well? What is their attitude towards indicators? Does it depend on being affiliated with certain groups?

## The Italian case

The press is in itself a variegated arena, characterized by diverse actors, practices, power dynamics, and national specificities as well (Brüggemann, 2014; Bucchi and Trench, 2014). To reduce the complexity, we decided to focus on a specific context, which we deem particularly relevant for our purposes: the Italian context.

The rationale for focusing on the use and representation of bibliometric indicators in the Italian press is twofold.

First, the Italian research evaluation system is, in Europe, the system that most heavily relies on bibliometric indicators (Baccini et al., 2019; Bonaccorsi, 2020). Bibliometrics is used to evaluate both institutions and individuals. At the institutional level, citation and publication output data feed the algorithms used in the periodic research assessment exercises, called VQR ("Valutazione della Qualità della Ricerca"), to evaluate the quality of the articles in the hard sciences, life sciences, and engineering (the so-called "bibliometric areas"). Peer-review is limited to the social sciences and humanities ("non bibliometric areas") and research outputs not covered by citation databases. At the individual level, researchers should meet at least 2 out of 3 bibliometric thresholds to be eligible for the National Scientific Habilitation (ASN), which is required in turn for applying to associate and full professorships. For the "bibliometric areas," these thresholds are calculated on three indicators: the number of journal articles, the number of citations, and the h-index. Both the VQR and the ASN are managed by the Agency for the Evaluation of the University and Research (ANVUR).[2] This complex research evaluation system was introduced in 2010 within a comprehensive reform of the governance and organization of Italian state universities (Law 240/2010).[3] The reform raised several contestations from the Italian academic community and, relevantly for the purposes of this study, was widely discussed in the Italian press (Commisso, 2013). The discussion of the reform constituted a significant channel for the introduction of bibliometric indicators to the Italian public opinion.

The second reason why the Italian context is particularly interesting is that it lacks a strong, indigenous bibliometric community. No research center comparable to the Dutch CWTS or the Canadian *Observatoire des sciences et des technologies* exists in Italy. ANVUR occasionally funds small research projects in bibliometrics but is not endowed with scientific staff on its own. The Italian bibliometric research groups are scattered among different institutions (universities, National Research Council) and academic disciplines, including economics, managerial engineering, and library and information science. In the bureaucratic classification of disciplinary areas used by the Italian Ministry of University and Research to classify Italian researchers, bibliometrics does not even count as an independent research area (CUN, n.d.). The lack of a strong bibliometric community is striking when it is compared with the considerable weight that bibliometrics has in the national research evaluation system. In this sense, the Italian case shows why it is crucial to improve Leydesdorff and colleagues' scheme by considering the weight of the different communities, as it may vary significantly between national contexts.

These two features of the Italian context, i.e., the centrality of bibliometrics in the research evaluation system on the one hand, and the lack of a strong community of professional bibliometricians on the other hand, have created the conditions for *newspapers* to become pivotal arenas for the discussion of bibliometric indicators in Italy. The press has constituted an important channel through which Italian researchers have been exposed to bibliometric indicators, parallel to that of the bureaucratic procedures of research evaluation implemented after the 2010 reform. In this sense, the representation of bibliometric indicators given in the newspapers, to which researchers themselves have contributed as authors of press articles or as interviewees, has been an important source of the "bibliometrics folk theories" (Aksnes and Rip, 2009) held by Italian researchers.

## Methods and gathered data

Four leading Italian daily newspapers were examined: *Corriere della Sera, La Repubblica, La Stampa*, and *Il Sole 24 Ore*. They are a standard reference for the study of science in the Italian press (Ampollini and Bucchi, 2020). The former three are the most widely circulated and prestigious Italian newspapers. *Il Sole 24 Ore* is a business newspaper owned by Confindustria, the Italian employers' federation, which traditionally reserves special attention to university and higher education affairs.

---

[2] https://www.anvur.it/en/agency/mission/ (accessed 17 November 2020).
[3] See Donina and colleagues (2015) for a reconstruction of the reforms of higher education governance in Italy from 1980s to present. A detailed criticism of the effects of 2010 reform on the Italian university is offered in (Viesti, 2016).The viewpoint of a member of ANVUR's board is provided in (Bonaccorsi, 2020).

The articles were gathered through the internal search engines of the newspapers' websites or online archives between July and August 2020.[4] Three sets of keywords were used to retrieve articles relevant to the research. A first set included the term "bibliometrics" ("bibliometria" in Italian) and related forms, both as a noun and an adjectival form (e.g., "bibliometrico"). The resulting articles are useful to investigate the representation of bibliometrics in general. The other two sets included the terms "h-index" and "Impact Factor", along with their most frequent Italian translations and variants, and were used to retrieve articles specifically mentioning these two indicators.[5]

All types of articles were included in the corpus, apart from duplicates and three advertisement articles. The result was a corpus of 583 articles divided into three sub-corpora: Bibliometrics, H-Index, and Impact Factor (Table 1). 45 articles are shared between two sub-corpora, 1 article appears in all the three (Figure 1). The first article in the corpus dates 1977 but it is an isolated case. All the other articles date from 1990 onwards.

|  | **Impact factor** | **Bibliometrics** | **H-index** | **Total (%)** |
|---|---|---|---|---|
| *La Repubblica* | 154 | 45 | 35 | **234 (40%)** |
| *Sole 24 Ore* | 83 | 84 | 22 | **189 (32%)** |
| *Corriere della Sera* | 71 | 26 | 0* | **97 (17%)** |
| *La Stampa* | 30 | 15 | 18 | **63 (11%)** |
| *Total (%)* | *338 (58%)* | *170 (29%)* | *75 (13%)* | ***583 (100%)**** |

*Table 1. Corpus composition, broken down by newspaper and sub-corpus. \* See Note 5. \*\* 536 distinct articles.*

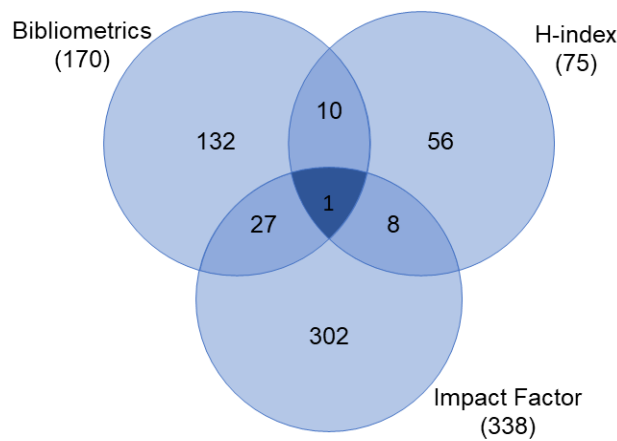

*Figure 1. Overlap between the three sub-corpora*

The articles were analyzed following a content analysis methodology (Bryman, 2015: 13). The dimensions to code the articles were derived from a careful and iterated reading of the articles themselves to individuate those that mostly capture the content of each sub-corpus (Elo and Kyngäs, 2008). The resulting dimensions were grouped into six main categories. The first category contains the meta-data of the article; the second some article's general features such as the profession of the author, the scientific field mentioned,[6] and the presence of interviews; the third the function of bibliometric indicators in the rhetorical structure of the article; the fourth the attitude of the article's author towards the indicators; the fifth the entity to which the indicator is attributed; and the sixth the attributes the indicator is associated with. In several categories, some sub-corpus-specific dimensions were individuated to code specific contents that appear only in certain sub-corpora. The complete list of dimensions is provided in Appendix 1.

---

[4] University **[omissis for anonymous review]** subscription was used to access the digital archive of *Sole 24 Ore*, a private account to access the digital archive of *Corriere della Sera*. *La Repubblica* and *La Stampa* websites can be accessed for free.

[5] It was not possible to retrieve the articles mentioning the h-index in *Corriere della Sera* due to some limitations in the newspaper archive's search engine (e.g., queries containing terms with hyphens are not allowed).

# Results

## Trends and research policy events

Figure 2 shows the distribution of corpus articles over time. Bars indicate the total number of articles per year, whereas the lines show the trend of each sub-corpus.[7] Most of the articles are concentrated in the decade 2010s when 66% of the articles appeared. 29% of the articles were published in the 2000s and only 4% in the 1990s. In particular, the incidence of articles remains low until the beginning of the century, it starts to increase during the first decade, and peaks in 2012. After 2012, the curve drops rapidly.[8]

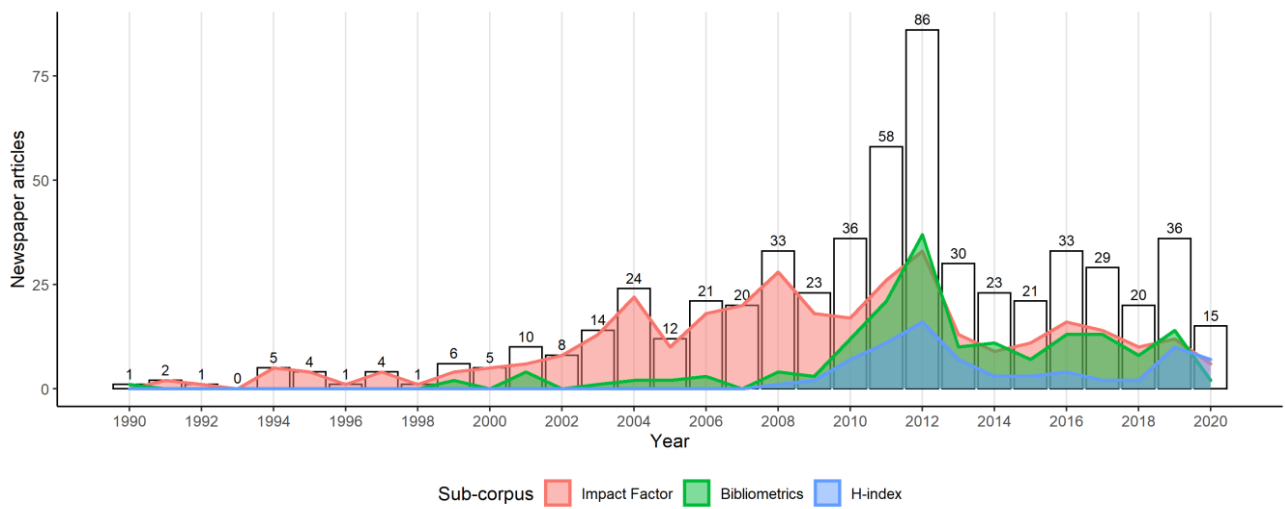

*Figure 2. Absolute number of articles over time (bars: overall, lines: broken down by sub-corpus).*

The three sub-corpora follow different trends. In general, the articles mentioning the IF show the highest frequency compared to the other two groups and, until 2010, they almost represent the totality of the articles. The trend of the IF sub-corpus includes several minor peaks in 2004 and 2008, and a major peak in 2012 (33 articles). After that, the incidence decreases, and the number of articles citing the IF remains close to the number of articles mentioning bibliometrics. The sub-corpus Bibliometrics counts a few articles until 2009. Then, it suddenly accelerates and reaches a peak in 2012 (37 articles) when it even overtakes the IF sub-corpus. After the peak, it slows down. As to the sub-corpus H-index, it should be noted that the relative indicator was created only in 2005. Therefore, it cannot be mentioned in the press before that year. Interestingly, however, it takes only 3 years for the h-index to consume the distance between the scientific community and the newspapers: the first article mentioning the h-index dates 2008. The H-index sub-corpus too displays a sudden acceleration, reaching again its peak in 2012 with 16 articles. Then, it decreases but remains a constant presence in the press.

---

[6] The classification of research fields was based on UNESCO's most recent classification (http://uis.unesco.org/).
[7] The oldest article in the corpus, that dates 1977, was removed from the plot.
[8] Data for 2020 are underestimated as the articles were collected between July and August 2020.

The temporal analysis points out the importance of 2012. Is there an event that can explain the peaks we found both overall and in each sub-corpus? The analysis of the content of 2012 articles reveals that most of them deal with the *first round of the National Scientific Habilitation (ASN)*, i.e., the procedure introduced with the university reform of 2010 that deeply affected the recruiting and career advancement mechanisms in Italian universities. As we saw in the Introduction, bibliometric indicators play a crucial role in the awarding of the ASN. The h-index is one of the three indicators used to evaluate applicants, which explains why it raised the attention of the press suddenly after the introduction of the reform. The years 2012-2014 show also a significant intensification of the engagement of academics in the press. Academics write 32 of the 86 articles of 2012 (37%) and for the next two years, the number of articles written by academics overcomes the number of articles written by journalists (the plot is shown in Appendix 2). The turbulent relationship with indicators and bibliometrics in these years is highlighted by the polarization of academics' attitude: of the 54 articles they write, 20 show a negative attitude (37%), 17 a positive attitude (31%), and only 4 (7%) adopt a balanced attitude where both positive and negative aspects of indicators are discussed. Note that in these years, the *individual dimension* of bibliometrics is particularly at stake, i.e., the fact that bibliometrics is used to evaluate individual scientists rather than institutions. In the sub-corpus Bibliometrics, which reflects the general climate towards metrics, individuals are the entities bibliometric measures are mostly attributed to in these years. The focus on the individual is furtherly reinforced by the attention paid to the h-index, which, in turn, reflects its role in the Habilitation. In these conditions, bibliometrics and performance indicators become a "personal affair" of Italian researchers, which may account for the polarization of attitudes noted above.

The key role of the first round of the ASN in explaining the 2012 peak may generate the false impression that articles in the corpus mostly deal with research policy events. Clearly, several articles directly address research policy, and they attest that the university reform was a crucial moment in the construction of the social discourse around indicators. However, the functions of bibliometrics indicators in the articles and the type of news where they appear are richer and more diverse.

## Overview of the main functions and contexts of use of bibliometrics

The content analysis of the corpus reveals that the IF and the h-index play a distinct set of *functions* in the rhetorical-argumentative structure of articles. The indicators are used by the articles' authors or by the persons interviewed to advance and motivate specific claims about a wide array of topics. As to the general occurrences of "bibliometrics" and related terms, collected in the sub-corpus Bibliometrics, it is better to talk of *contexts of use* rather than functions, as bibliometrics-related terms are less integrated into well-delimited rhetorical structures.

Notably, the IF and the h-index share almost the same set of functions. First, they are both presented as means to *promote meritocracy* in the university. Indicators are introduced as objective measures that can fix what is perceived as a distinctive trait of the Italian university, namely the clientelism in academic recruitment. As we will better see in the next section, the indicators are introduced as "justice devices" to highlight the injustices committed by the selection committee. Second, indicators are used to *praise the scientific performance* of individual scientists. Scientific success is thus quantified and associated with indicators, backing up with numbers the judgment of excellence made in the press article. Note that also the IF is used to commend individuals, despite, from a technical point of view, it is a journal-level indicator. A third, related function is that of *defending the scientific reputation* of a researcher in a controversy. In these articles, frequently written by researchers themselves, the ability to publish in "journals with high IF" or directly showcasing the values of bibliometric indicators allows the authors to defend their scientific status when their reputation is attacked.

Indicators, however, are not only attributed to individuals. One of their most common function in the corpus is that of *measuring and then commenting on the scientific performance of countries, universities, and hospitals*. The performance of the Italian research system is in particular assessed with indicators, to compare it to European competitors or to show that performance is not matched by adequate funding. As to universities, indicators are closely associated with university rankings and, in local editions of newspapers, the performance of the local university is praised (or dispraised) by mentioning, among other things, their bibliometric performance. Lastly, *hospitals and the IF* are closely associated in the press: the "IF of a hospital" is routinely reported by journalists and frequently used in articles published in the local edition of a newspaper to praise the local excellence. Again, from a strictly technical point of view, attributing an IF to an organization diverges from the standard definition of the indicator. However, the "IF points"[9] are included among the standard metrics used in Italy to evaluate the performance of hospitals and it is not uncommon for a hospital to showcase their IF on its web-site[10] (Horenberg et al., 2020). In this regard, journalists simply follow the common practices of hospital performance assessment in Italy.

Besides these performance assessment-related functions, the IF plays also a distinctive *epistemic* role, not played by the h-index. The IF is used to *certify the scientific credibility of science news*, as a sort of "quality seals" that supports the reliability of scientific discovery. In this sense, it is an important component of the construction of scientific facts that takes place in the media (Bucchi, 1998).

However, not all articles share the confidence in indicators that characterize the functions mentioned so far. In fact, there is a group of articles where indicators play the role of *targets of criticism*.

In the next sections, each of these functions will be detailed and the most characteristic themes associated with them analyzed. From a quantitative point of view, however, their incidence is shown in Table 2. Advancing meritocracy is the most represented function of the IF, whereas the h-index is mostly used to praise individuals. Both the indicators are targets of criticism in a consistent quota of articles.

| Function | Impact Factor | | | H-index | | |
|---|---|---|---|---|---|---|
| | Articles | % | Rank | Articles | % | Rank |
| Meritocracy | 90 | 26,6% | 1 | 15 | 20,0% | ▼2 |
| Hospitals | 79 | 23,4% | 2 | - | - | - |
| Criticism | 40 | 11,8% | 3 | 13 | 17,3% | ●3 |
| Universities | 31 | 9,2% | 4 | 8 | 10,7% | ▼5 |
| Credibility | 24 | 7,1% | 5 | - | - | - |
| Praise | 22 | 6,5% | 6 | 21 | 28,0% | ▲1 |
| Controversy | 20 | 5,9% | 7 | 4 | 5,3% | ●7 |
| Other | 17 | 5,0% | 8 | 6 | 8,0% | ▲6 |
| Country (Italy) | 15 | 4,4% | 9 | 8 | 10,7% | ▲4 |

---

[9] They correspond to the sum of the IF of the journals where the physicians of a certain hospital publish.
[10] E.g.: https://research.hsr.it/en/highlights.html (accessed 17 November 2020).

| | Total | 338 | 100,0% | | 75 | 100,0% | |

*Table 2. Functions of the IF and the h-index in the respective sub-corpora*

In the Bibliometrics sub-corpus, the contexts of use of bibliometrics-related terms partly overlap with the functions found in the two other sub-corpora. Bibliometric terms are used to *certify the excellence* of individuals and institutions, to comment on the *scientific performance of Italy*, and are mentioned when *rankings of universities* are presented. However, they mainly appear in the context of the *discussion of the research evaluation system*. Bibliometrics is mentioned again as an instrument to promote meritocracy in academic recruitment, as it offers "objective measures" that can repair the subjectivity of the committee's judgment. Bibliometric terms also occur in the description of the procedures of the Italian research evaluation system (VQR, ASN).[11] Notably, most of the articles where bibliometrics is the *target of criticism* are about the evaluation system.

Table 3 reports their absolute and relative frequencies of each context in the sub-corpus Bibliometrics. Again, it is clear the preeminence of the discussion over research evaluation, with the first two contexts accounting alone for more than half of the articles.

| Rank | Context of use | Articles | % |
|---|---|---|---|
| 1 | Research Evaluation | 64 | 37,6% |
| 2 | Criticism | 29 | 17,1% |
| 3 | Rankings | 19 | 11,2% |
| 4 | Perfunctory | 16 | 9,4% |
| 5 | Country (Italy) | 15 | 8,8% |
| 6 | Excellence[12] | 14 | 8,2% |
| 7 | Other | 12 | 7,1% |
| 8 | Credibility | 1 | 0,6% |
| | *Total* | *170* | *100,0%* |

*Table 3. Incidence of bibliometrics' contexts of use in the sub-corpus Bibliometrics.*

## Indicators as justice devices and the meritocracy frame

The earliest articles where the IF occur, dating back to the middle of the 1990s, share a common topic: they are all about scandals in university recruiting procedures. In Italy, professors and researchers are appointed by open competitions ("concorsi pubblici"), in which a committee of peers is in charge of selecting the best candidate. The law provides that the choice of the candidate must be based only on scientific merit, independently of academic interests or irrelevant qualities of the candidates, such as gender or political affiliation. Nonetheless, "old boys" academic circles, localism, cronyism, and other forms of clientelism are perceived by many as an endemic problem of the Italian academy (Battiston, 2002; Morano Foadi, 2006). Leaving aside the question whether the perception is true or not, it should be noted that these issues are deeply integrated into the *narration* of university affairs unfolding in the Italian press (De Nicolao, 2012). Indeed, they are also the context that hosts the earliest occurrence of the Impact Factor in the Italian newspapers. Bibliometric indicators find their way to the press in news about scandals in the university, being presented as "*justice devices*" that can remedy injustices.

At the beginning of the 1990s, Italy faces a period of nationwide intense judicial investigation into political corruption, known as "tangentopoli" (after "tangente", literally "bribe"), resulting in the demise of many traditional political parties. Amid this climate, a series of scandals in the recruitment of professors reaches the attention of the media, and it is explicitly paralleled to the cases of political corruption (the term "concorsopoli" is coined in assonance with "tangentopoli")[13]. In the articles in our corpus covering these scandals, the candidates that were unjustly rejected by the committees find a new means to demonstrate to the public the injustice they suffered: the IF. In these years, in Medicine it become a common practice to sum the IFs of the journals where a medical doctor has published **[REF!]**, obtaining an IF-based metric for individuals. In this way, candidates that lost in open competitions had at their disposal a simple indicator by which their scientific performance could be measured and, most importantly, *easily compared* to that of the winners. Moreover, the availability of a simple number gives the public the impression to enter the realm of scientific judgment, opening the possibility to judge whether the appointed researcher was really scientifically superior to the competitors or not. The presence of the IF, therefore, helps the candidates and the journalists writing about open competitions in university to frame the narration in terms of injustice and scandals, also echoing the general climate of the tangentopoli period. In this sense, the IF functions as a "justice device" able to exhibit to the public the unfairness of recruitment in university, as the following quote shows:

> In the invalidated Clinical Oncology competition, two winners had an IF of 36 and 24 points, against the 359 points of Professor Robin Foà, who was rejected. In that of Internal Medicine (still *sub judice*), to the candidate Pandolfi, the absolute best with 130 IF points, was preferred his colleague Aliberti, who gathers only 8 points.[14]

---
[11] When bibliometrics occurred only in the phrases "bibliometric areas" and "non-bibliometric areas", the context of use of that article was classified as perfunctory.
[12] Without reference to benchmarking or rankings.

In the 1990s, most of the articles mentioning the IF are about specific competitions (see Figure 3). This use continues in the next decade. In 2004, another big scandal in Cardiology, involving several medical doctors in the whole country, receives significant coverage in the media. One-third of the 24 articles from this year in the corpus are devoted to the scandal. This time, the telephones of the professors were even tapped and the articles report excerpts from their conversation, where the ways to manipulate differences in the IF of the candidates are openly discussed. Again, the perception of injustice is reinforced in the reader by how journalists report and comment on the IF:

> The impact factor [of the selected candidates] (the score based on scientific publications, very important for the committee's judgment) was about ten times lower than Picano's (to be sure, even someone on the commission had a lower score than that of the candidate).[15]

The comparison, made possible by the use of numbers and conveyed with simple mathematical operations («ten times lower»), allows even to *measure* the magnitude of the injustice.

In the following years, the articles that mention the IF in the contexts of specific competitions continue to appear until the university reform of 2010, but with lower rates on the total (see Figure 3). This happens because the indicator assumes new functions in the press but also because the enthusiasm for indicators as justice devices seems to be waning in recent years.

---

[13] Marco Panara: "Concorsopoli, polveriera dell'università". *La Repubblica,* 11 July 1995.
[14] Riccardo Chiaberge: "Università, altri concorsi a rischio". *Il Corriere della Sera*, 31 October 1994. The original quotes in Italian are reported in Appendix 3.
[15] Michele Bocci: "Medicina, l'inchiesta si allarga". *La Repubblica,* 20 may 2004.

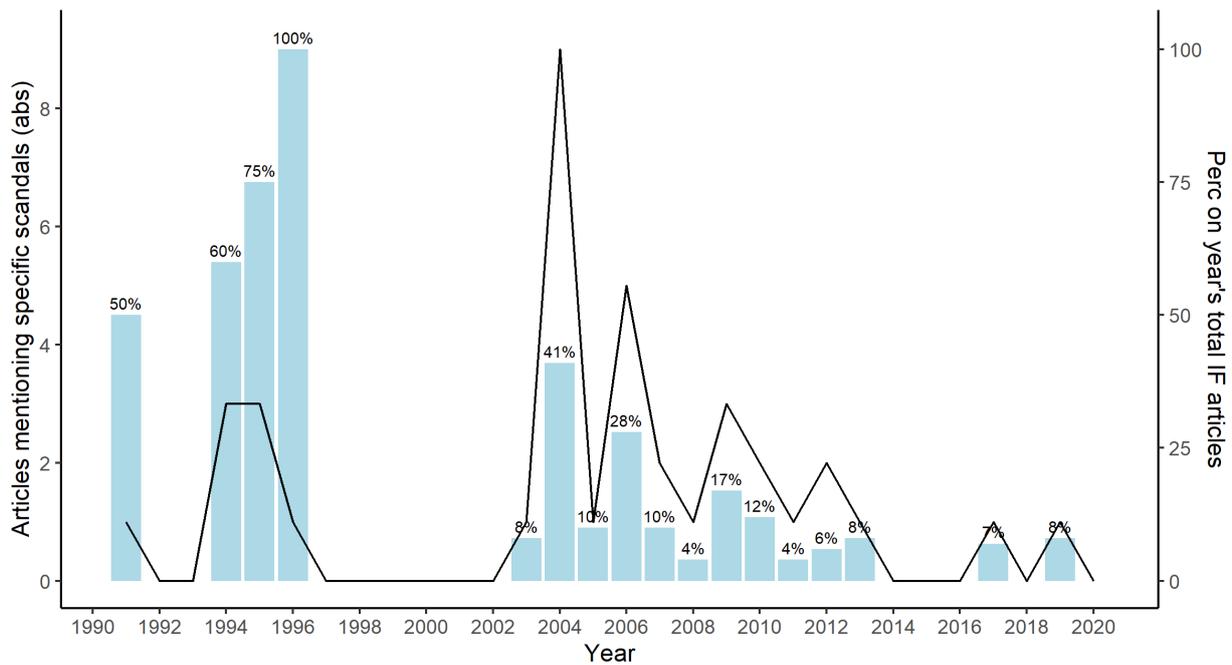

*Figure 3. Articles mentioning specific scandals (bars: percentage on year's total IF articles, line: absolute number of articles).*

The h-index is less mentioned in news about individual competitions. However, it appears in one article that may be considered the epitome of the "justice device" function. [16] In this article, the journalist reports in the very first lines the values of *five* bibliometrics indicators for the two candidates to a professorship, h-index included. Then, she invites the reader to judge who of them is the best. Note that no definition of the indicators is provided, neither the content of the research of the two candidates is described. It is only said that they compete for a position in Agricultural Sciences in the same university. After this bibliometric showcase, the journalist writes that the committee chose the candidate with the lower scores. The clash between the committee's judgment and the bibliometric numbers openly offered to the reader generates a two-fold effect. On the one hand, it induces a clear perception that injustice has occurred. On the other hand, that bibliometric measures *alone* are enough to judge scientific merit. Furthermore, it suggests a normative consequence: if academic recruitment would be based on "objective indicators", meritocracy would be promoted in the university. Indicators are thus presented as *drivers of meritocracy*.

The association between indicators and meritocracy and the idea that indicators can be a cure for the familism and cronyism of the Italian university, are one of the main themes of the entire corpus. Figure 4 shows its incidence over time, both in overall and broken down by sub-corpus.

Note how most of the articles (96 articles, 72%) are concentrated between 2004 (the year of the big Cardiology scandal) and 2012, the year of ASN first round. The peak in 2008 coincides with debates around the reform of the university the Italian government had announced and that become law at the end of 2010. Interestingly, bibliometrics and bibliometric indicators are presented as drivers of meritocracy mostly in the period *before the university reform* and when the debates on the national scientific habilitation take place. After this period of press excitement, the theme's weight significantly falls.

---

[16] Flavia Amabile: "Dov'è la meritocrazia a Milano?". *La Stampa,* 28 may 2012.

The temporal analysis shows how indicators were integrated into a meritocracy-centered frame long before they were officially enrolled in the Italian research evaluation system. In fact, they circulated in the press with this function for twenty years before the system was set up.

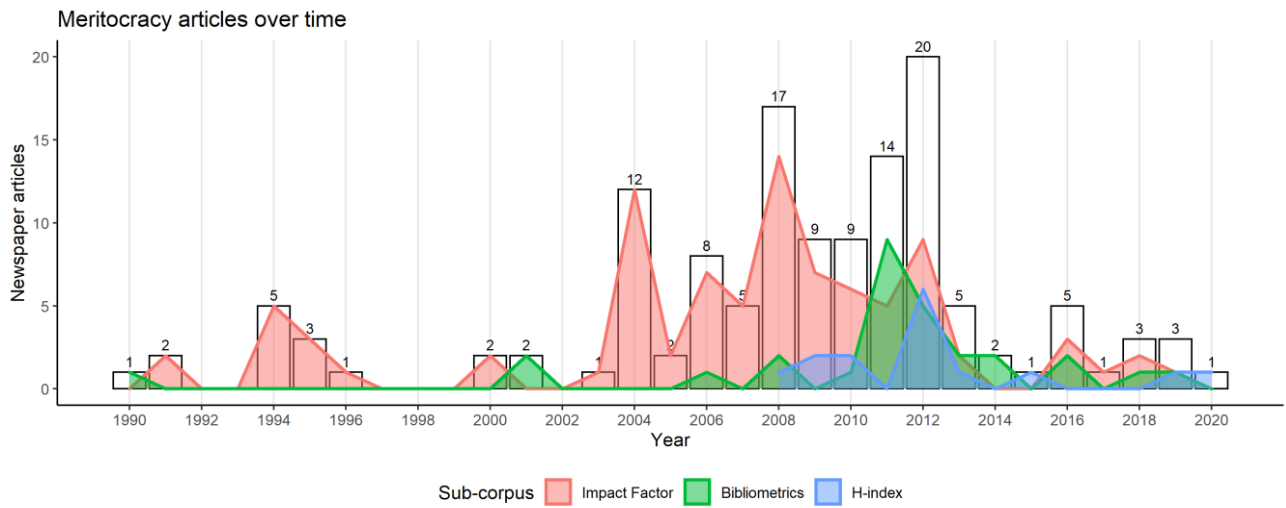

*Figure 4. Articles belonging to the meritocracy frame over time (bars: overall, lines: broken down by sub-corpus).*

## Indicators as praising devices

Both the IF and the h-index are used as "praising devices" in the corpus, i.e., to celebrate the scientific excellence of scientific actors, mostly individual scientists but also hospitals and even single publications. Journalists and scientific journalists are the most frequent type of authors that use the IF and the h-index in this way, writing 21 of the 43 articles characterized by this function. The reference to a scientific performance metric in an article communicates to the public the idea that there are objective methods to award scientific excellence, immune to subjective biases. Especially when individuals are celebrated with indicators, the metric is frequently backed up by its specific numerical value (13 articles on 28 celebrating individuals report the numerical value of the indicators), reinforcing the idea that scientific excellence is bestowed based on objective numbers and not on the subjective impression of the journalist or elusive notions such as prestige. That the mere mention of metrics and numbers is deemed sufficient to exhibit scientific excellence is shown by the fact that 74% of the articles belonging to this function lack any definition of the indicators mentioned or provide too vague definitions, and 91% of the articles do not mention or discuss limitations of bibliometric indicators. The lack of definitions and caveats contribute to generating a "halo of objectivity" around bibliometric measures.

Almost in all cases, the scientists that are celebrated by metrics are medical doctors or life scientists, showing how indicators are deeply entrenched not only in the intra-specialist epistemic culture of these fields but also in their communicative practices with external audiences. Bibliometric indicators have been so integrated into the public image of medical scientists, that some of them are routinely introduced by a bibliometric-centered "epithet": this is the case of Alberto Mantovani, a famous Italian immunologist that frequently intervenes in the press, whose standard introduction in the newspapers includes phrases such as «one of the top 100 most cited immunologists»[17], «the scientist that bibliometric analyses indicate as the most productive and cited in the world scientific literature»[18], or «a scientist with an h-index that places him among the most important immunologists in the world»[19]. He is also permanently mentioned in the articles reporting the ranking of the most cited Italian scientists, closely associating his public image with performance indicators.

---

[17] "Mantovani direttore di Humanitas". *Il Sole 24 Ore*, 5 November 2005.
[18] Marco Pivato: "Ecco i soldati anti-tumore". *La Stampa,* 12 July 2012.

Consistently with its being an individual-level metric, the h-index most frequent function in the corpus is praising the scientific performance of individual scientists: 28% of the articles in the sub-corpus are characterized by this function. Interestingly, the use of the IF as a celebrating device starts to be less sporadic from 2009 onwards, in coincidence with the advent of the h-index in the Italian press, as if the presence of the other praising indicators has spilled over the IF functional characterization. However, the IF serves to celebrate scientific actors only in 7% of the relative sub-corpus.

In the sub-corpus Bibliometrics, bibliometrics appears in articles focusing on the "excellence" of scientific actors in 14 articles (8.2%). Rather than individuals, however, the focus of these articles is the excellent performance of universities. Not surprisingly, academics are the authors of 43% of these articles.

## The Impact Factor and the construction of scientific facts in the press

The use of the IF as a "credibility seal" for science news is recent: of the 24 articles where the indicator plays this function, 22 appeared after 2008. In these articles, the IF, correctly attributed to journals, is presented as a warrant of the scientific quality of the venue of publication, and hence, of the credibility or relevance of the science news reported:

> The research, just published in the prestigious journal Cerebral Cortex (Impact Factor 8.3), presents the first data in the world of a specific genetic variation related to the visual impairment associated with dyslexia.[20]

Both journalists, scientist journalists, and academics use the indicator in this way, sharing a common positive attitude towards it. 18 articles mention the title of the scientific journal and 9 even report the numerical value of the IF. However, only 7 articles include some definition of the indicator that may help the reader to interpret its meaning. Of these, only 2 are correct definitions. In the rest of the articles, the simple mention of the "high value" of the IF is enough to provide scientific credentials, as this quote shows:

> Now the consecration of the new approach has arrived, with the publication of an article in the "Cancer Journal for Clinicians", the journal with the greatest "impact factor" in the world.[21]

Note that no article where the IF has this function mentions the limitations of the indicator, that hence is presented as a completely "transparent" quality seal, easily interpretable and uncontested.

Scientific news in the medical and health sciences are the most covered in this group of articles, accounting for 63% of the total. Again, this shows how tightly the IF is integrated into the communicative practices of these scientific areas. News in the natural sciences (astrophysics, biology, and physics) follows at distance with 4 articles, whereas only 2 articles report science news coming from psychology.

---

[19] Piero Bianucci: "Covid 2020: santa alleanza tra vaccino e sistema immunitario". *La Stampa,* 15 June 2020.
[20] Rosalba Miceli; "Vulnerabilità genetica". *La Stampa,* 10 October 2014.
[21] Sara Ricotta Voza: "Chi salva il cuore con i biomarcatori ha più possibilità di vincere il tumore". *La Stampa*, 7 July 2016.

Interestingly, the IF is used as a seal of scientific quality even in articles that report news from areas outside official science, such as homeopathy. These articles, written by a homeopath medical doctor, report a study that allegedly showed the efficacy of homeopathic therapies in mice.[22] The fact that the study was published in a journal «with a good impact factor» is used to buttress the scientific credentials of the alleged discovery. In a follow-up article, the title of the study, the name of the journal, and the value of the IF to the third decimal position «according to Thomson Reuters' Journal Citation Report» are meticulously reported, along with the DOI and a link to the study.

By the same token, the IF can be used also to contest the scientific credentials of a discovery and deconstruct its credibility. Once again in an article on homeopathy, the reliability of a study published in *Scientific Reports* is contested mentioning the low IF of the journal of publication. The author of the article (a philosopher and historian of medicine), writes:

> If we consider the *impact factor*, which measures the scientific value of a journal with the number of citations received, that of *Nature* is 40, whereas *Scientific Reports* slightly exceeds 4.[23]

These cases show that the IF is a powerful tool in the communication of science news in the press. The IF contributes to the construction and deconstruction of scientific claims by providing a seal of scientific quality that can be strategically conferred or withhold to scientific discoveries.

A somehow related function of bibliometric indicators is that of *defending the scientific reputation of individual scientists or institutions in controversies* (24 articles in the corpus, 4%). This function starts to appear at the beginning of the century and the controversies mainly occur in the field of medicine, often in connection with the management of hospitals. The IF and other bibliometric indicators are mobilized to respond to accuses of bad management or low scientific performance.

Among the articles where the reputation of individuals is at stake, the most curious case is that of a chemist accused of sexism because he used a naked model on the graphical abstract of a publication on coconut milk. Interviewed by the journalist, he defends his scientific respectability by citing his h-index and comparing it to that of Albert Einstein:

> I travel the world and I am used to being evaluated only for my scientific skills. I have an H-Index of 60, Einstein had 100. It means that as a researcher I reached Everest.[24]

## The advent of the h-index in the Italian press and the role of "amateur bibliometrics"

It takes only three years for the h-index, created in 2005, to reach the Italian press. Its advent in the news sheds light on the peculiar role that *amateur bibliometrics* (Gläser and Laudel, 2007) may play in the communication of bibliometric indicators to the public. The carrier of the h-index in the Italian press was a ranking of scientists produced by an Italian researcher who collected data as a private initiative, without any institutional support. The ranking included the Italian scientists with an h-index higher than 30 (calculated on Google Scholar) and was populated by automatic data retrieving methods, the software Publish or Perish, and by collecting personal recommendations. It was named "Top Italian Scientists" (TIS) and published online by the association Via-Academy (Virtual Italian Academy). It is still active today.[25] Figure 5 shows the key role of this initiative in the early days of the h-index life in the Italian press: in 2010 and 2011, most of the articles that mentioned the indicator were in fact about the TIS. Overall, 1 out of 4 of the articles in the H-index sub-corpus cites the ranking. The style of these articles is quite homogeneous: scientists included in the ranking are praised for their scientific excellence, the strength of scientific areas is assessed based on the number of "top Italian scientists" they comprise,[26] and the fact that many of these scientists work abroad is lamented as a sign of the decline of Italy.[27]

---

[22] Alberto Magnetti: "Anche i topolini sono sensibili alle cure omeopatiche". *La Stampa*, 15 May 2010; Alberto Magentti: "Medicinali omeopatici: dimostrata la riproducibilità in laboratorio". *La Stampa,* 19 April 2012.
[23] Chiara Lalli: "È un inganno spacciare l'omeopatia per scienza". *Il Corriere della Sera,* 11 November 2018.
[24] Paolo Berizzi: "Ma quale sessismo, la modella svestita fa capire la scienza". *La Repubblica*, 27 March 2014.

Despite the centrality of the h-index in the generation of the ranking that is the topic of these articles, 6 articles in the group (32%) lack any definition of the indicator, 10 (53%) provide a vague or imprecise definition, and only 2 report the correct definition.[28] Limitations, such as the fact that the h-index is not field-normalized and hence the values of scientists working in different fields cannot be easily compared, are mentioned in 9 articles out of 19.

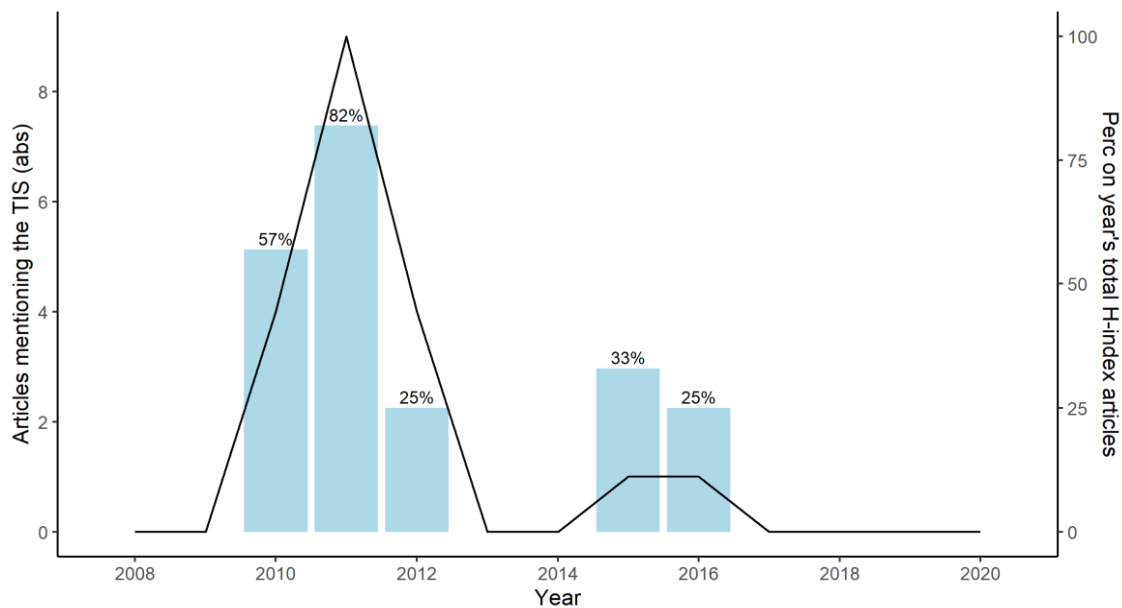

*Figure 5. Articles mentioning the Top Italian Scientists ranking (bars: percentage on year's total H-index articles, line: absolute number of articles).*

Interestingly, the close association between the h-index and amateur bibliometrics practices emerges also in the very first article mentioning the indicator in the corpus. This article is especially relevant because it was written by Roberto Battiston, head of the Italian National Institute for Nuclear Physics (INFN), member of the Council of Experts in Research Policy (an advisory group of the Italian Ministry of Research), and expert evaluator for ANVUR. Thus, the author is an important figure in the Italian research policy environment. In this article, he commends the software Publish or Perish as the fastest and easiest means to measure the level of meritocracy in the Italian university:

> In a few minutes, it is possible to create the ranking of merit of an entire university department, having the satisfaction of understanding, based on perfectible but objective factors, how one is evaluated by the reference community. At least, you can include the H-index in your resume or compare it with that of the commissioners of your next open competition.[29]

---

[25] https://topitalianscientists.org/home (accessed November 17, 2020).
[26] Flavia Amabile: "Gli astrofisici rimontano". *La Stampa*, 29 May 2011.
[27] Flavia Amabile: "Dove sono gli informatici italiani?". *La Stampa,* 13 December 2010. « Six out of 100 Italian talents in the world deal with computers. none of them work in Italy. In a ranking of the Via-Academy the crisis of the Italian IT industry»
[28] To be sure, neither the website of the ranking provides the technical definition of the h-index.

The reference to the "perfectibility" of indicators slightly softens the three key messages: that bibliometrics is objective, that it promotes meritocracy, and that no specific bibliometric expertise is needed to calculate and interpret the metrics. Amateur bibliometrics and the meritocracy frame are thus deeply connected.

## Bibliometrics and rankings

The Top Italian Scientists ranking is not the only ranking mentioned in the corpus. In fact, in the sub-corpus Bibliometrics, more than 1 in 10 articles mentions various types of rankings (Table 3). International university rankings start to appear in the corpus from 2009 onwards. The three most mentioned are the ARWU ranking, the Times Higher Education ranking, and the Leiden ranking (entirely based on bibliometrics). The domestic rankings of universities produced by ANVUR based on the VQR exercise and by the agency's predecessor, the CIVR (Comitato di indirizzo per la valutazione della ricerca) appear in the corpus as well. They are specially covered in the pages of *La Repubblica*.

What is more interesting, however, is that the relationship between rankings, bibliometrics, and newspapers in Italy *predates* the advent of international rankings and the launch of the research evaluation system. In fact, a group of articles appeared in *Corriere della Sera* between 2003 and 2006 publicize the results of a *ranking of the Italian hospitals* that the newspaper produced for several years in collaboration with the Mario Negri Institute for Pharmaceutical Research, a nonprofit research institute dedicated to clinical and biomedical research. Notably, these rankings were entirely based on the bibliometric analysis of the scientific output of hospitals. They aimed explicitly at measuring the quality of the research done in the hospitals, which was deemed to affect and thus also represent the quality of the cures provided. The IF played a central role in this early assessment exercise, which, it is worth noting, was a private initiative of the newspaper. It did not receive support from the Italian government and the Ministry of Health even criticized the ranking as a source of misinformation for patients.[30] In this regard, the publication of *Corriere*'s ranking of hospitals provoked the appearance of the first articles where the IF is *contested*. It is interesting to note that one of the most informed critical articles in this period was written by an association of oncological patients. It presented a detailed criticism of the IF reporting the main arguments of the bibliometric literature and even quoted Eugene Garfield's cautions on it.[31]

*Corriere*'s ranking of hospitals shows that newspapers do not act only as passive conduits for bibliometric knowledge produced by other social actors. They can also be *active producers* of bibliometrics and hence participate "firsthand" in the social construction of the indicators' meaning through their editorial choices. In fact, *Corriere*'s ranking initiative was crucial in the early circulation of bibliometric indicators in Italy and contributed to their legitimization as reliable measures of scientific performance. At the same time, this initiative provoked also the first *critical reactions*. By contrast, in the earliest occurrences of the IF, when the indicator played the role of justice device, it was never contested.

---

[29] Roberto Battison: "Cogito ergo pubblico". *Il Sole 24 Ore,* 25 September 2008.
[30] "I malati (e i medici) disorientati". *Il Corriere della Sera,* 23 May 2004.
[31] "All'Alleanza non piacciono le nostre classifiche del cancro". *Il Corriere della Sera*, 4 July 2004.

## Critical voices in the press

Most of the ways in which indicators and bibliometrics are used in the press imply a certain degree of confidence in the capacity of bibliometric measures to capture scientific performance or even excellence. However, there is a group of articles where this assumption is questioned, and the indicators become a target of criticism. In the sub-corpora IF, H-index, and Bibliometrics, they amount, respectively, to 12%, 17%, and 17% of the articles (see Table 2 and Table 3 above).

The critical arguments found in these articles can be grouped into three main families. The most common revolves around the idea that the use of indicators in the evaluation system would have dramatic *epistemic consequences* on the advancement of science. Indicators would advantage those approaches and schools that belong to the "scientific mainstream". Hence, the pluralism of the scientific inquiry would be seriously damaged, with potentially revolutionary ideas being suppressed by conformism,[32] and fashionable topics being artificially inflated.[33] A closely related argument is that bibliometrics-based evaluation would favor certain schools of thought at the expense of others in the same discipline, independently of their scientific merit but only in function of their relationship with the Anglo-American world that produces the indicators. This line of argument is especially evident in a series of articles devoted to disputes in the discipline of economics, in which the use of the IF in evaluation is criticized as it would favor American neo-liberal schools at the expense of Italian schools.[34]

A second family of criticism argues that citations would measure the mere popularity of a topic in the scientific community, rather than scientific excellence. In this regard, they are compared to the audience's measures of television shows.[35]

The last family of criticism insists on the fact that the adoption of indicators will induce strategic behavior in researchers, such as the practice of dividing a scientific study into multiple publications ("salami slicing")[36] and, more generally, privilege quantity over quality.[37]

In addition to charting specific criticisms, we assessed also the general *attitude towards indicators and bibliometrics* that authors express in their articles. In 85 articles (15% of the total), this attitude is negative or openly critical (Figure 6). These critical articles are mostly concentrated in the Bibliometrics sub-corpus, in which 24% of the articles are denoted by a negative attitude. Compared to the other sub-corpora, moreover, this sub-corpus is characterized by a higher incidence of "balanced" articles, where both strengths and shortcomings of the indicators are presented. The IF and especially the h-index, on the other hand, polarize more the attitude of the authors.

---

[32] Carlo Galli: "Perché bisogna poter valutare la ricerca". *La Repubblica,* 5 August 2010. Gianfranco Rebora: "Spazio alla valutazione 'duale'". *Il Sole 24 Ore,* 29 August 2011. Vittorio Lingiardi: "Impact Factor con giudizio". *Il Sole 24 Ore*, 30 June 2013.
[33] Riccardo Viale and Loet Leydesdorff: "Il paradosso della ricerca più citata". *Il Sole 24 Ore,* 8 June 2003. Piero Bianucci: "Bolle scientifiche e bolle finanziarie". *La Stampa,* 1 July 2015.
[34] Franco Locatelli: "La contesa che divide gli economisti", *Il Sole 24 Ore,* 12 November 2004. Saverino Salvemini: "Non imitiamo gli USA negli errori". *Il Corriere della Sera,* 11 December 2006. Federico Fubini: "La lettera degli economisti: 'Studi, meno conformismo'". *Il Corriere della Sera*, 3 March 2009.
[35] Cesare Segre: "La quantità non è un criterio per valutare la qualità". *Il Corriere della Sera*, 6 February 2009. Giuseppe Galasso: "Inaffidabile la pagella per le riviste". *Il Corriere della Sera,* 12 May 2011. Tullio Gregory: "La retorica dell'inglese per tutti". *Il Corriere della Sera*, 7 March 2012.
[36] Roberto Casati: "Paradossi della valutazione". *Il Sole 24 Ore,* 30 October 2011.
[37] Salvo D'Agostino: "Pubblicare o morire, così va la scienza oggi?". *Il Sole 24 Ore,* 1 April 2012. Michela Marzano: "Pubblica o muori, quel nuovo sistema che spegne il sapere". *La Repubblica,* 17 February 2013.

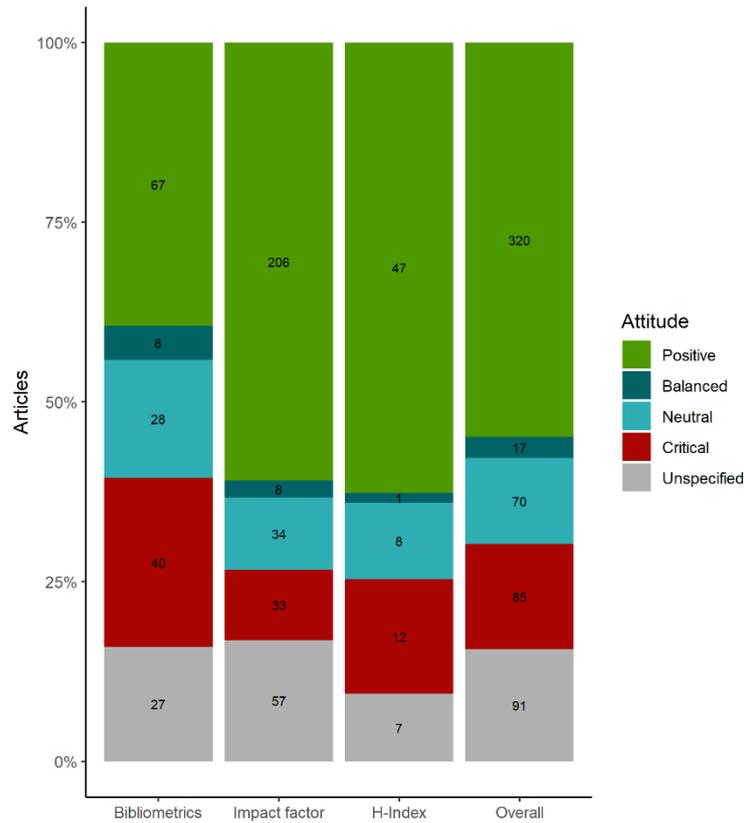

Figure 6. Author attitudes, broken down by sub-corpus and in overall

The temporal analysis shows that both the positive and the critical attitudes peak in 2012, when, as we have seen above, the discussion on the National Scientific Habilitation is at the center of the stage (Figure 7). Balance articles surge as well, even if they remain by far the less represented group.

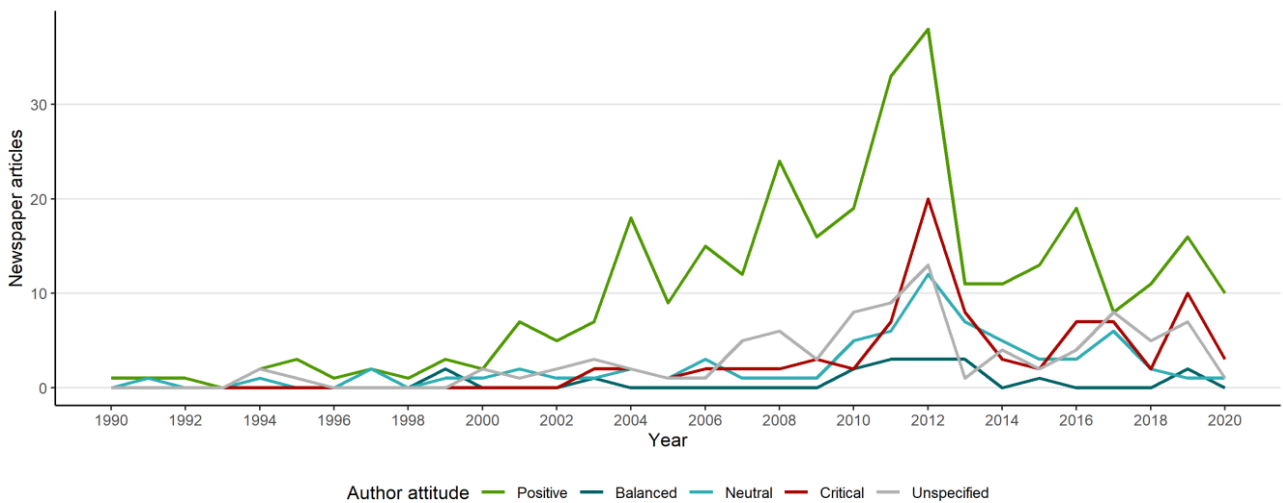

Figure 7. Author attitudes over time

Who are the authors that show the most negative attitude towards bibliometrics and bibliometric indicators? The data show that they are mainly *academics working in the* *humanities* (Figure 8). Overall, humanities scholars express a critical attitude in half of the articles they write. To compare, academics in Medicine express a negative attitude in less than 10% of their articles and a positive attitude in 68% of their articles. In the years around the first round of the National Scientific Habilitation (2012-2014), humanists write 39% of the articles, showing a negative attitude in two-thirds of them. However, the criticisms they issue do not concern the technical limitations of bibliometric methodologies: humanities scholars do not mention any specifical technical shortcoming in 35 of their articles (69%) and do not provide any definition of the h-index and the IF in 65% of the articles where they mention these indicators. Rather, their main criticism is that *bibliometrics does not work for the humanities* and hence should not be used to evaluate their areas. In addition, they criticize research assessment from a wider point of view: for instance, two eminent Italian philosophers argue that the very idea of *measuring* research performance should be rejected as the epitome of neo-liberalism applied to higher education.[38]

Interestingly, the key contribution of humanists to the social discourse around bibliometrics, i.e., the specificity of the humanities, is shared by other academic actors and journalists as well, showing that it has been successfully incorporated in the debate. Of the 64 articles in the sub-corpus Bibliometrics that mention the humanities, 32 are written but journalists, academic professors in the sciences, rectors, and even science policy officials. Thus, the specificity of the humanities is successfully communicated to the public and, indeed, seems to be adequately integrated into the Italian research policy as well, where the humanities are considered "non-bibliometric areas".

---

[38] Roberto Esposito: "Valutazione ossessione". *La Repubblica,* 17 February 2013. Pier Aldo Rovatti: "Quanto è difficile valutare i docenti". *La Repubblica,* 17 February 2013.

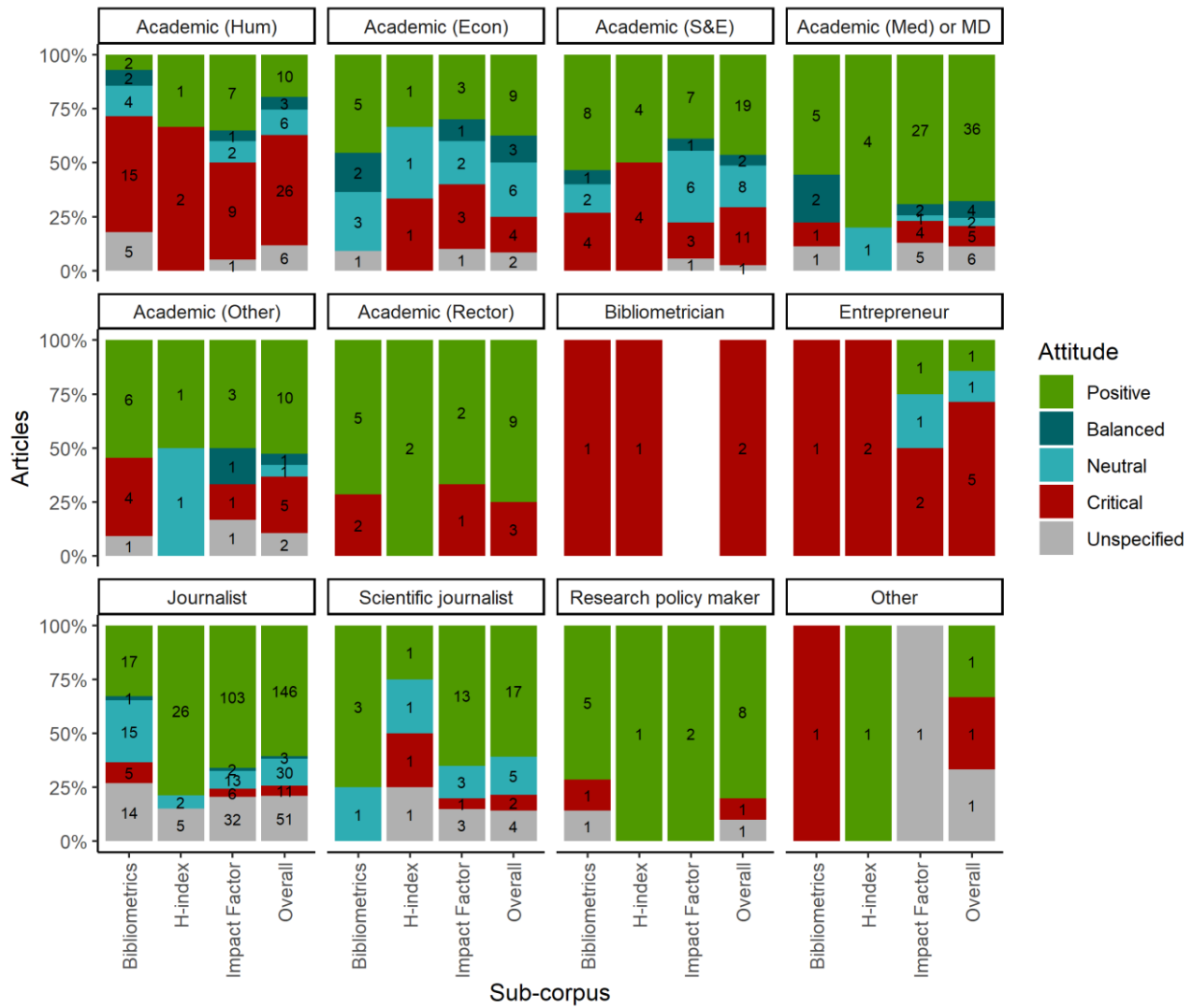

*Figure 8. Author attitudes towards bibliometrics, broken down by groups of authors and sub-corpora*

## Groups of actors in the press

As Figure 8 shows, journalists write most of the articles, but they are not the only group represented in the corpus. Table 4 reports the incidence of the different types of actors, showing both the number of articles they author and the number of articles in which they are interviewed. These two modes of engagement in the social construction of indicators reflect differences in the role of groups: some groups intervene actively by writing articles, whereas others bring their viewpoints and interest through interviews. Belonging to a group that is frequently interviewed by journalists arguably reflects a perception of expertise or authoritativeness on the topic. On the other hand, writing firsthand attests both the intention to address the public opinion without the mediation of journalists and the capacity to obtain room in prestigious newspapers.

The actor types can be grouped into four main macro-categories: journalists, academics, medical doctors, and research policy officials. Journalists include both generalist and journalist specialized in science topics. Academics are divided by their field, excluding academics in Medicine, which are included in the category of medical doctors, as they are presented in articles and interviews more as physicians than as academics. Lastly, research policy officials comprehend Ministry of Research and ANVUR officials. Again, these persons are frequently academics, but in articles and interviews, their association with research policy structures is preeminent.

Besides generalist journalists, who write 41% of the articles, medical doctors are both the most prolific group of authors (9% of the articles) and the category most interviewed (45% of the interviews). Interestingly, in the ranking of productivity, they are closely followed by scholars in the humanities, who, however, are seldomly interviewed (they write 9% of the articles but appear only in 4% of the interviews). Academics in science and engineering are less prolific authors but are more frequently interviewed (12% of the interviews). Economists are characterized by a profile similar to that of humanists: they directly engage as authors but are not proportionally chosen for interviews. Academic rectors and research policy officials, by contrast, rarely engage firsthand in the press, but they are frequently interviewed. Interviews with research policy officials are more than double the articles they write firsthand. Most of the academics are Italian and are based in Italian institutions, but it is not uncommon for Italians affiliated with universities in other countries to intervene.[39] Lastly, actors outside the university sector have a scarce weight in the corpus: entrepreneurs write only 7 articles in the entire corpus.

Only 2 articles are written by professional bibliometricians: one (co-)authored by Loet Leydesdorff and another by Giovanni Abramo.[40] Interestingly, both articles are highly critical of the misuse of bibliometric indicators. Leydesdorff warns of the epistemic dangers of the use of indicators, which may promote regressive research programmes, whereas Abramo harshly criticizes the Top Italian Scientists ranking based on the h-index and the role of the indicator in the National Scientific Habilitation.

| Author Category | Articles (%) | Rank | Interviews (%) | Rank |
|---|---|---|---|---|
| *Journalist (total)* | *269 (46%)* | | - | |
|     Journalists | 241 (41%) | 1 | - | |
|     Scientific journalists | 28 (5%) | 5 | - | |
| *Academics (total, excluding medical doctors)* | *147 (25%)* | | *72 (35%)* | |
|     Academics (Humanities) | 51 (9%) | 3 | 8 (4%) | ▼6 |
|     Academics (Science and Engineering) | 41 (7%) | 4 | 25 (12%) | ●4 |
|     Academics (Economist) | 24 (4%) | 6 | 7 (3%) | ▼7 |
|     Academics (Other) | 19 (3%) | 7 | 2 (1%) | ▼8 |
|     Academics (Rector or similar) | 12 (2%) | 8 | 30 (15%) | ▲2 |
|     (Bibliometricians) | 2 (0%) | 12 | 1 (0%) | ▲9 |
| Academics (Medicine) or Medical Doctors | 53 (9%) | 2 | 91 (45%) | ▲1 |
| Research policy officials | 10 (2%) | 9 | 24 (12%) | ▲3 |
| Entrepreneurs | 7 (1%) | 10 | 1 (0%) | ▲9 |
| Other | 3 (1%) | 11 | 15 (7%) | ▲5 |
| [Unknown] | 92 (16%) | | 0 (0%) | |
| **Total** | **583** | | **204** | |

*Table 4. Incidence of types of actors in articles and interviews*

---

[39] E.g., Roberto Casati: "Sotto tiro l'agenzia francese". *Il Sole 24 Ore,* 14 October 2012. Casati is currently research director at the French National Research Council (CNRS).
[40] Riccardo Viale and Loet Leydesdorff: "Il paradosso della ricerca più citata". *Il Sole 24 Ore,* 8 June 2003. Giovanni Abramo: "Le trappole che ostacolano le donne nella scienza". *La Stampa*, 10 June 2020.

## The translation of bibliometric expert knowledge in the press

If we term "bibliometric expert knowledge" the ensemble of intellectual contents that are discussed in bibliometric journals (see the Introduction), it is possible to assess how specific items of this intellectual repertoire are represented in our corpus. In this way, we can trace the modifications expert bibliometrics undergoes when it is translated into the medium of the press.

The most basic items relevant to our corpus are the *definitions* of IF and h-index. In the bibliometrics literature, they are standard because both indicators received a precise definition by their creators, Eugene Garfield and Jorge Hirsch. In our corpus, these technical definitions struggle to penetrate the articles. As Figure 9 shows, 49% and 58% of the articles do not report *any definition* for the h-index and the IF, respectively. When a definition is provided, 10% of the articles in the IF sub-corpus and 7% of the articles in the H-index sub-corpus provide a wrong one. In both sub-corpora, the quota of articles providing the correct definition is only 5%.[41]

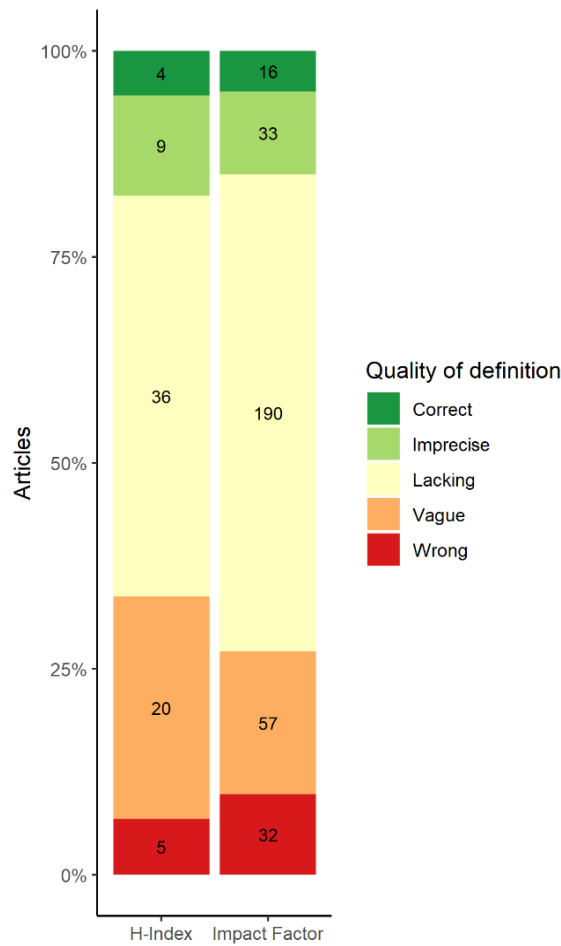

*Figure 9. Quality of bibliometric indicators' definition provided in the articles*

The significant absence of definitions that emerges from this analysis can be intended as the sign of a wider process, which is of particular interest for tracing the social construction of the indicators in the newspapers, namely the *dropping of indicators' modalities*. Modalities are statements that contextualize and limit the validity of bibliometric indicators and, from a certain point of view, are the distinctive trait of bibliometric expertise (Gläser and Laudel, 2007;

---

[41] By correct, we mean a definition that contains all the main elements of the standard definition of the indicators, not a definition expressed in formal terms.

Latour and Woolgar, 1986). Bibliometrics, as an intellectual endeavor, mainly deals with understanding the correct interpretation of bibliometric data, which means essentially circumscribing their validity. The combined effect of removing definitions and dropping modalities induce, by contrast, a charge of *objectivity, soundness, and consensus* on bibliometric indicators that lack in the bibliometrics expert literature (Gläser and Laudel, 2007: 105–108).

The analysis of our corpus, summarized in Figure 10, reveals that modalities, in the form of limitations and caveats to the interpretation of indicators, are cited in only a limited portion of the articles. Limitations are explicitly explained and discussed in only 8% of the corpus, whereas 9% of the articles simply mention their existence without further elaboration. Interestingly, in the sub-corpus IF, the quota of articles that lack any reference to limitations reaches 90%, showing that the "solidity" of the IF is taken almost for granted in the corpus.

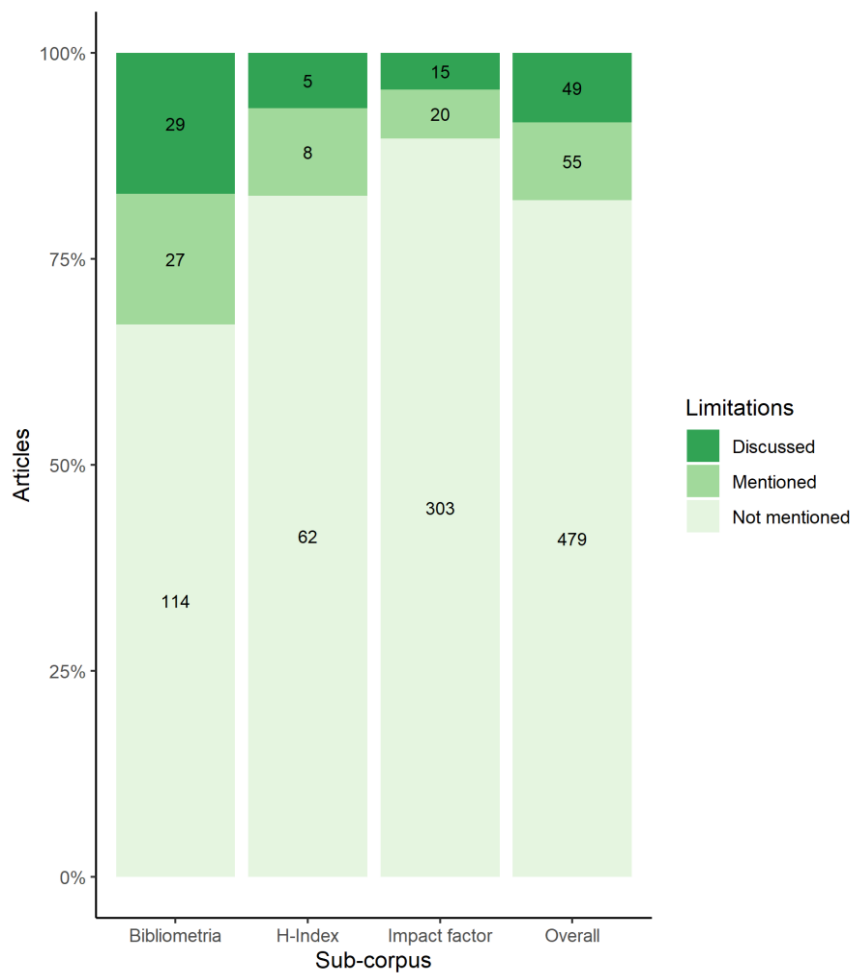

*Figure 10. Distribution of articles mentioning limitations, broken down by sub-corpus and in overall*

Besides definitions and modalities, another key item of bibliometric expert knowledge is the correct attribution of indicators to certain entities. As we saw in the Introduction, the IF is a journal-level metric, whereas the h-index is an individual-level metric.[42] In our corpus, the attribution is, by contrast, widely more varied. As Figure 11 shows, both indicators and bibliometrics-related terms are attributed to the entire spectrum of scientific actors, from the single publication to the country. Notably, individual researchers are associated with indicators in all three sub-corpora. This focus on individuals is coherent with the main functions that indicators play in the press, which work at their best when the indicators are attributed to individuals. For instance, when the IF is used to demonstrate an injustice in open competitions, attributing it to journals would weaken the rhetorical strength of the article and the claim to justice of rejected candidates. In these contexts, technical correctness gives way to rhetorical needs. On the other hand, the individual focus contributes to transforming indicators into a "personal affair" for Italian researchers, as they become crucially associated with the assessment of *individual* performance.

---

[42] To be precise, the calculation procedure the indicators are based on can be applied to any entity: it is possible, for instance, to calculate the h-index of a journal. However, both the indicators were *originally* targeted on specific entities of the scientific process, the journal and the individual researcher, respectively.

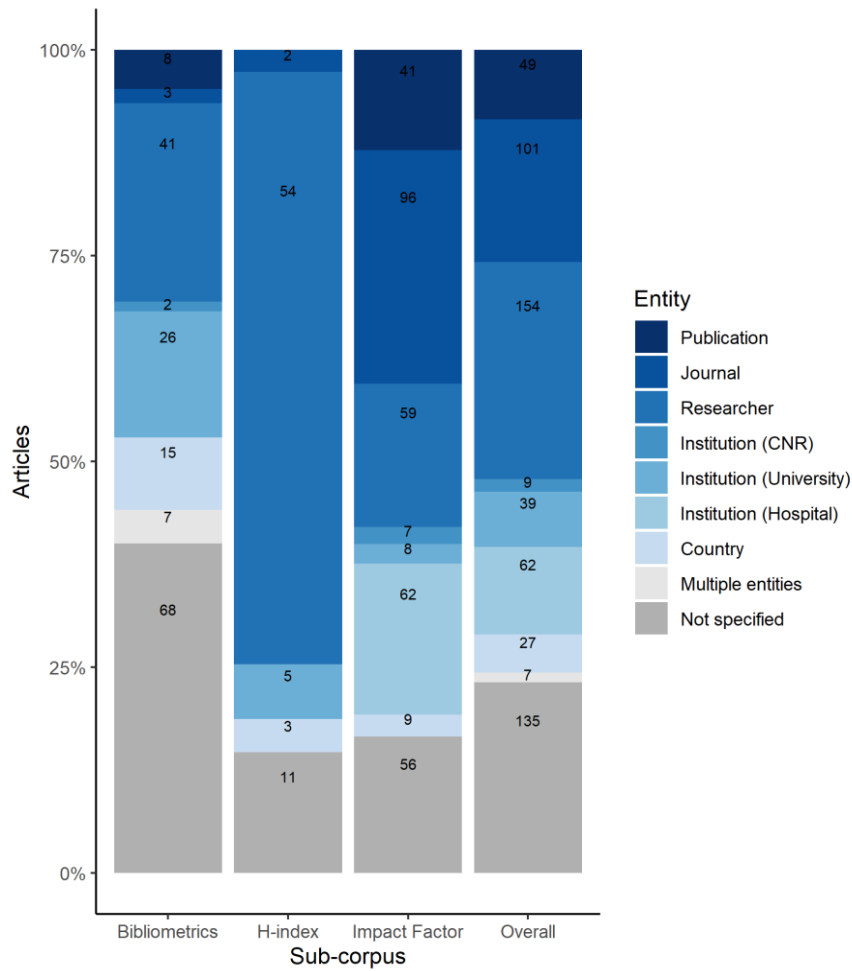

*Figure 11. Entity attribution of indicators and bibliometrics, broken down by sub-corpus and in overall.*

So far, we have seen items of the bibliometric expert knowledge that are deeply transformed when they enter the press. There is, however, an item, which we already mentioned in the previous paragraph, that arrives almost untouched in the articles: the idea that bibliometric methods should be used with extreme caution in the evaluation of the humanities. The bibliometric community has reached a certain consensus that the scarce coverage of bibliometric databases in these areas, the variety of publication outlets, the relevance of local languages, and the different function of references in humanities publications hinder the straightforward application of methods such as citation analysis to the humanities, and to a lesser extent, to the social sciences (Hammarfelt, 2017; Nederhof, 2006). As we showed above, this item of bibliometric knowledge has been well integrated into the Italian press and is well communicated to the public opinion.

To complete the charting of bibliometrics knowledge items in our corpus, we tracked which other bibliometric indicators are mentioned in the articles and counted their occurrences. It results that the indicators mentioned in more than 10 articles are only two: the number of citations (mentioned in 53 articles) and the number of publications (26 articles). More sophisticated indicators, such as the percentage of top publications or the field-weighted citation index appear only in 3 and 2 articles, respectively. The number of self-citations is cited in two articles. Therefore, the *advanced* bibliometric knowledge that, according to bibliometricians, is necessary to adequately and fairly assess research performance (Bornmann and Marx, 2014; Waltman, 2016), is almost absent from our corpus. In this regard, the considerable weight of the h-index, an indicator that is almost unanimously condemned by the bibliometric community (Waltman and van Eck, 2012), is revealing.

Besides the import of expert bibliometrics knowledge, can we trace the contours of a bibliometric theory specific to the press, a "folk bibliometrics theory" of the press (Aksnes and Rip, 2009)? The term theory is probably excessive. Nonetheless, there are two attributes that authors in our corpus commonly associate with bibliometrics and indicators and that are not imported from the specialist community of bibliometricians. The first is the charge of *objectivity* we discussed above. The term "objective" appears explicitly in association with bibliometrics only in 11 articles, concentrated in the sub-corpus Bibliometrics. However, the dropping of the modalities, the lack of definition, and the rhetorical functions, all convey the epistemic solidity of indicators, which is reinforced furtherly by the use of metaphors in articles about rankings, such as "photograph", "grades" and "report cards of professors". For instance, a journalist presents the first research evaluation exercise (VQR) in the following terms:

> Now under examination are the professors and their research: what they write, what they publish, their studies in the last seven years. We will know how research in Italy is doing and the grades that the departments and universities will get in the different disciplines. A wide-angle photograph focused on the scholarly production in the sciences and humanities. It is known as an "evaluation system" and will deliver a "report card" through which the best universities will be rewarded in the funding allocation.[43]

The second attribute is *internationality*. Both the two indicators and bibliometrics-related terms are presented in 21% of the articles as "widely diffused abroad" or as "best international practices".[44] Notably, this association intensifies between 2004 and 2012, when it occurs in 29% of the articles, then it slowly decreases. The contraposition between Italy and the "abroad", which frequently coincides with the Anglo-American world, is common in the articles of the meritocracy function, where bibliometric methods are frequently described as best practices that warrant meritocracy outside Italy. Internationally is hence a further piece in the mosaic of the meritocracy frame.

# Bibliometrics in the daily press: final remarks

Scholars in science and technology studies and bibliometricians are increasingly revealing the *performative nature* of bibliometric indicators (Müller and de Rijcke, 2017; Wouters, 2014). Far from being neutral measures, they are changing the social and epistemic structures of contemporary science through their incorporation into research evaluation mechanisms (Baccini et al., 2019; Rijcke et al., 2016; Wouters, 2018). At the same time, scholars have highlighted the *plasticity* of bibliometric indicators' meaning. Besides their technical definitions, the bibliometric indicators are endowed with *social meanings* that are constantly reframed in different contexts of practice (Gläser and Laudel, 2007; Hammarfelt and Rushforth, 2017; Leydesdorff et al., 2016).

---

[43] Laura Montanari: "Le pagelle dei professori". *La Repubblica,* 16 November 2011.
[44] For instance, Luigi Frati, rector of Rome university La Sapienza, says in an interview: «Research evaluation will be based on the *impact factor*» and the journalists comment in brackets: «an *international* scientific method for research evaluation» (emphasis added). Francesco Di Frischia and Simona De Santis: "Più ricerca, meno sprechi. Così cambio la Sapienza". *Il Corriere della Sera,* 3 February 2010.

In this study, we focused on a social arena that, so far, has not yet been investigated in this stream of research, namely the arena of the *daily press*. The content analysis of a corpus of 583 articles that appeared in four leading Italian newspapers in the last three decades allowed us to cast new light on how the discourse on indicators is socially constructed by both scientific and non-scientific actors. Three main results emerge from our material.

Firstly, generalist newspapers can be privileged channels to affect the public opinion in matters of research policy and hence promote policy agendas. The ways in which bibliometrics and bibliometric indicators were used in the Italian press, albeit diverse, converged in creating a *positive representation* of bibliometric indicators that was *conducive* to the establishment of a specific type of research evaluation system, where indicators play a key role. Indicators such as the IF were depicted as a *transparent and easy remedy* for the chronic problems of the Italian university system *long before* the reform was realized, and the bibliometrics-centered evaluation system implemented. This representation was mainly constructed when indicators played the role of justice devices but it was implicitly promoted also in those articles where they functioned as praising devices, seals of scientific quality, and clearly, reliable metrics for ranking individuals and institutions. In this regard, our analyses show how the roots of indicators-centered research policies may lie in processes of meaning construction that span over decades. Furthermore, they highlight that the newspapers play a crucial role in countries like Italy, which lack an indigenous bibliometric community able to take control of the discourse and influence policy makers.

Secondly, our material points out the preeminence that *medicine* and the *healthcare system* had in the construction of the social meaning of indicators in Italy. Most of the science news marked by the "IF seal" is about medicine or health topics, confirming early findings in the communication of science in the media, according to which medical topics are the most covered in the news (Bucchi and Mazzolini, 2003). More interestingly, however, it points out how bibliometric indicators are deeply entrenched not only in the intra-specialist epistemic practices of medical scientists but also in their *communicative strategies* with external audiences. As we have seen, certain medical doctors even received bibliometrics-based epithets in the press.

Lastly, the collected articles show how the existence of specific bibliometrics expertise and certified bibliometric knowledge struggles to be recognized in the press, at least in the Italian context. This is attested, on the one hand, by the key role that *amateur initiatives* played in the circulation of bibliometrics in Italy and, on the other hand, by the *scarce penetration of items of bibliometric expert knowledge*, such as the definitions and limitations of bibliometric indicators. These results confirm the difficulty of bibliometrics as a field to achieve exclusive professional jurisdiction on bibliometric indicators (Jappe et al., 2018).

An important topic for the study of bibliometrics in the press will be to assess whether these results are specific to the Italian context or they can be generalized to other countries. However, we hope to have shown that the press should not be overlooked in the analysis of the social construction of bibliometric indicators.

# Appendix 1: List of coding dimensions

| Category | Dimension | Sub-corpus | | |
| --- | --- | --- | --- | --- |
| | | **Bibliometrics** | **H-Index** | **Impact Factor** |
| 1 - Article metadata | 1a - ID | ● | ● | ● |
| | 1b - Sub-corpus | ● | ● | ● |
| | 1c – Newspaper | ● | ● | ● |
| | 1d – Title | ● | ● | ● |
| | 1e - Newspaper section | ● | ● | ● |
| | 1f - Newspaper edition | ● | ● | ● |
| | 1g – Page | ● | ● | ● |
| | 1h – Date | ● | ● | ● |
| | 1i – Author | ● | ● | ● |
| 2 - General features | 2a - Author profession | ● | ● | ● |
| | 2b - Scientific field | ● | ● | ● |
| | 2c - Presence of interviews | ● | ● | ● |
| | 2d - Interviewee(s) identity | ● | ● | ● |
| 3- Function / Context of use | 3a – Function | | ● | ● |
| | 3b - Mention the "TIS" ranking | | ● | |
| | 3c - Context of use | ● | | |
| | 3d – Mention of the humanities | ● | | |
| 4 – Attitude | 4a - Author attitude | ● | ● | ● |
| | 4b - Interviewee(s) attitude | ● | ● | ● |
| 5 – Attribution | 5a - Entity of attribution | ● | ● | ● |
| | 5b - Presence of numerical value | | ● | ● |
| | 5c - Mention of specific journal(s) | | | ● |
| 6 -Attributes | 6a - Quality of indicator's definition | | ● | ● |
| | 6b - Mention of limitations | ● | ● | ● |
| | 6c - Mention of the commercial nature of the IF | | | ● |
| | 6d - Association with internationality | ● | | ● |
| | 6e - Association with objectivity | ● | | |
| | 6f - Bibliometric indicators mentioned | ● | | |
| | 6g - University rankings mentioned | ● | | |
| | **Total** | **23** | **21** | **23** |

# Appendix 2: Further plots and tables

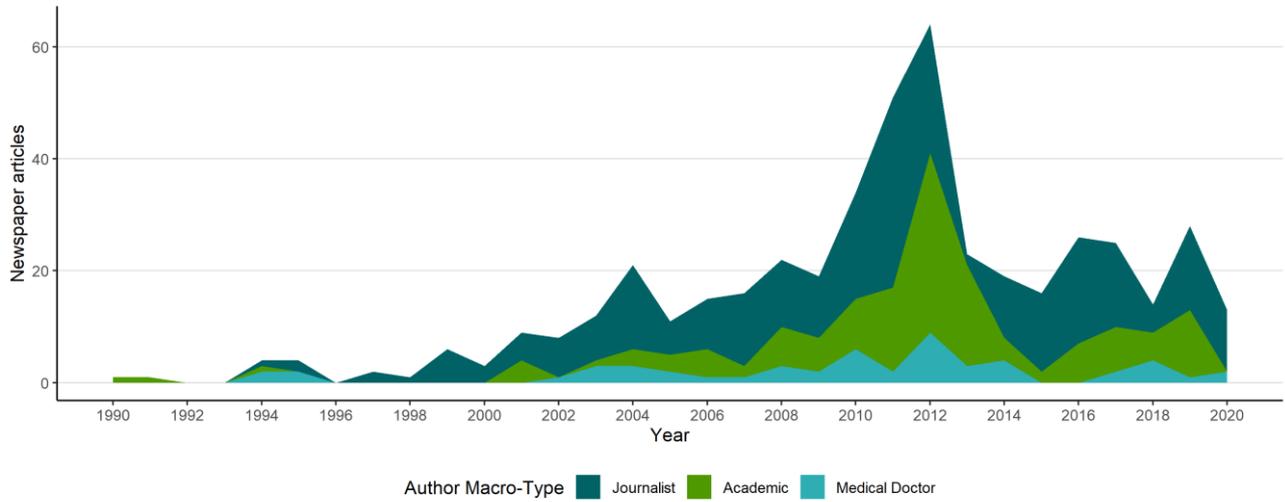

*Author Macro-Type over time*

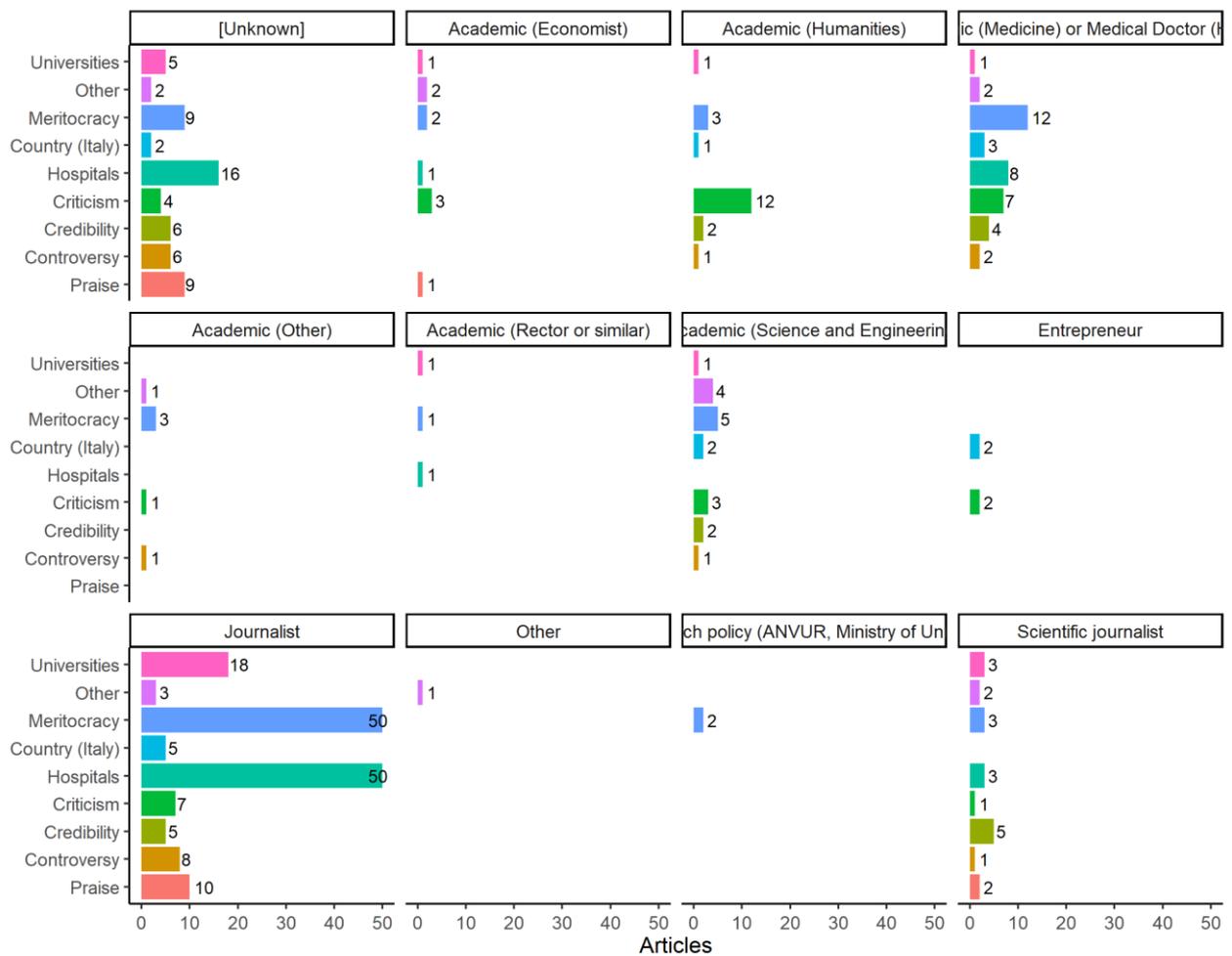

*IF and H-index' functions by author group*

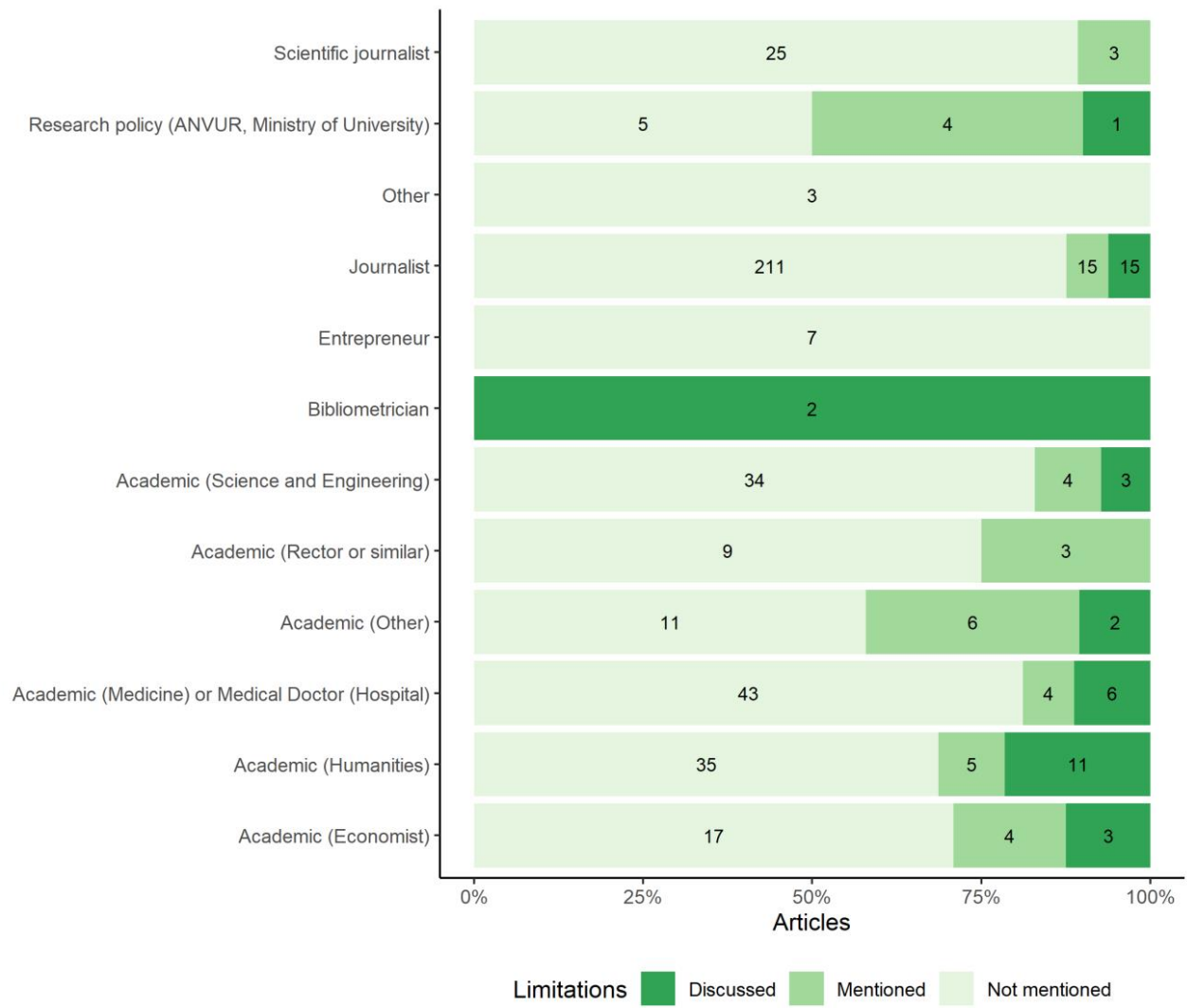

*Limitations by author group*

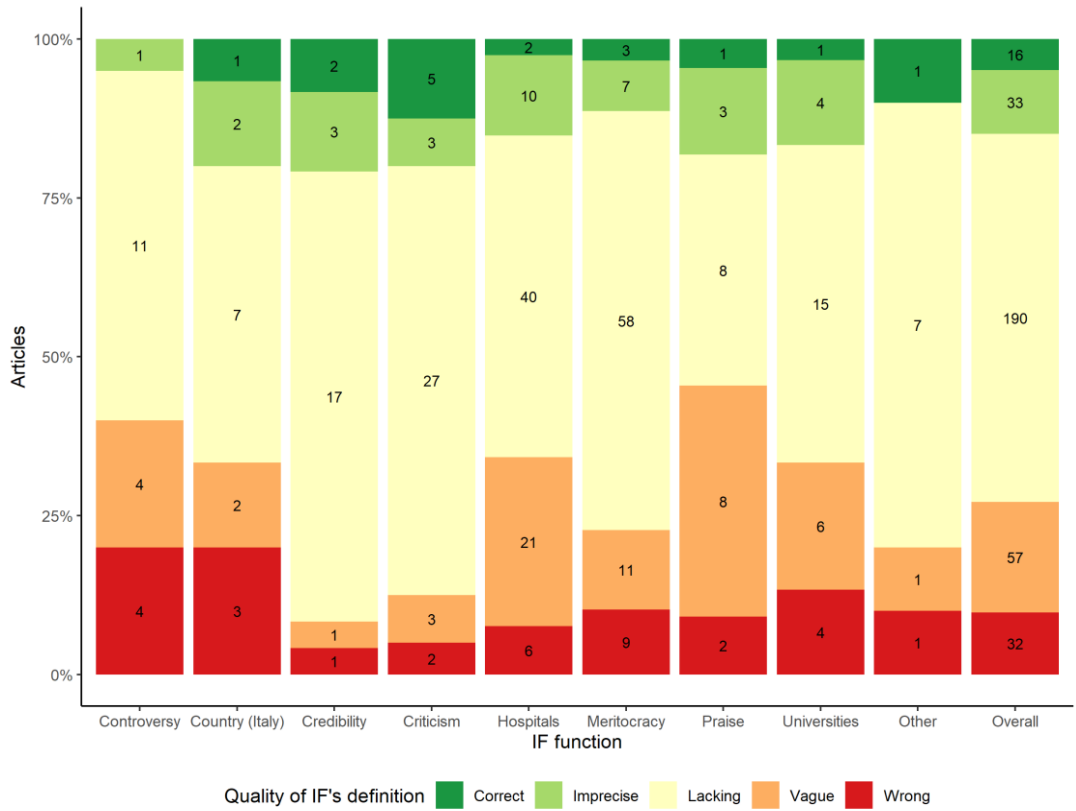

*Quality of the IF's definition by IF function*

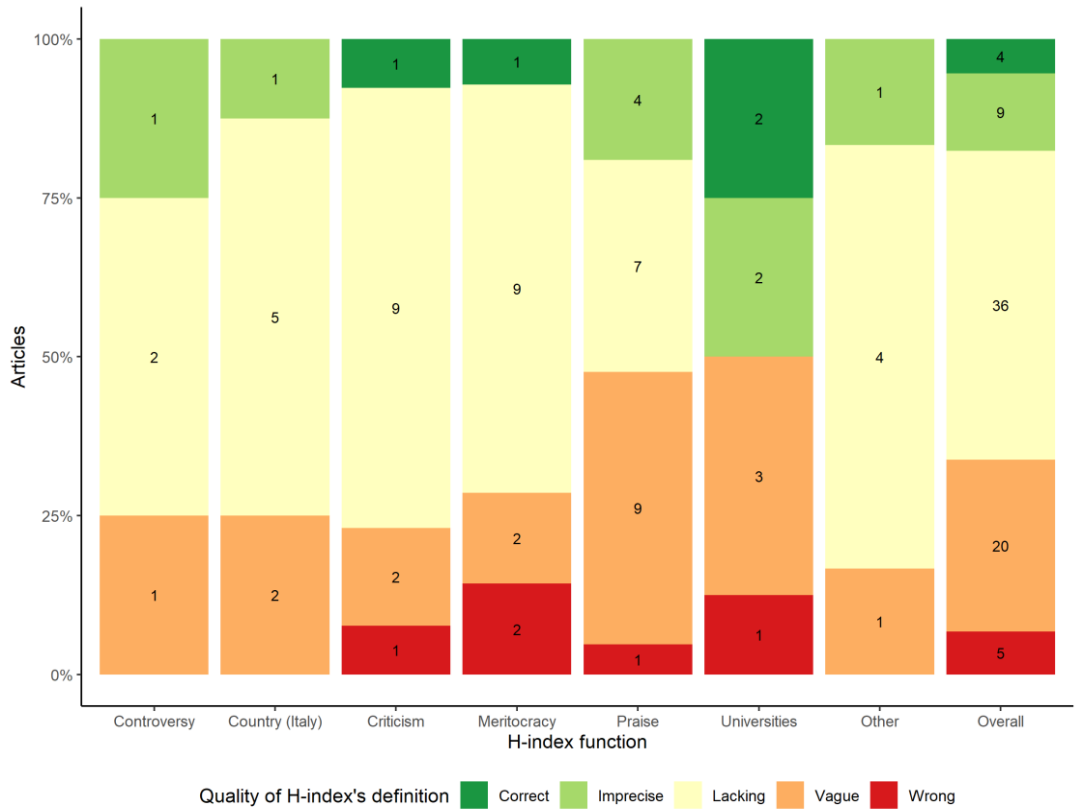

*Quality of h-index's definition by function*

| Scientific field | Bibliometrics | H-Index | Impact factor | Overall |
|---|---|---|---|---|
| **Healthcare System** | **1 (1%)** | **0 (0%)** | **110 (33%)** | **111 (19%)** |
| **Medical and Health Sciences** | **12 (7%)** | **18 (24%)** | **78 (23%)** | **108 (19%)** |
| Medicine | 12 | 16 | 78 | 106 |
| Pharmaceutics | 0 | 1 | 0 | 1 |
| Neurosciences | 0 | 1 | 0 | 1 |
| **Social Sciences** | **12 (7%)** | **2 (3%)** | **15 (4%)** | **29 (5%)** |
| Economics | 6 | 2 | 9 | 17 |
| Law | 2 | 0 | 2 | 4 |
| Psychology | 1 | 0 | 3 | 4 |
| Sociology | 2 | 0 | 1 | 3 |
| Political science | 1 | 0 | 0 | 1 |
| **Natural sciences** | **6 (4%)** | **4 (5%)** | **13 (4%)** | **23 (4%)** |
| Physics | 1 | 0 | 7 | 8 |
| Biology | 1 | 1 | 3 | 5 |
| Astrophysics | 0 | 1 | 2 | 3 |
| Chemistry | 1 | 1 | 1 | 3 |
| Mathematics | 2 | 0 | 0 | 2 |
| Geology | 1 | 1 | 0 | 2 |
| **Humanities** | **14 (8%)** | **1 (1%)** | **1 (0%)** | **16 (3%)** |
| Humanities (no specific area) | 13 | 1 | 0 | 14 |
| Philosophy | 1 | 0 | 0 | 1 |
| History | 0 | 0 | 1 | 1 |
| **Agricultural Sciences** | **3 (2%)** | **1 (1%)** | **4 (1%)** | **8 (1%)** |
| **Engineering and Technology** | **2 (1%)** | **3 (4%)** | **3 (1%)** | **8 (1%)** |
| Engineering | 0 | 2 | 2 | 4 |
| Architecture | 1 | 0 | 1 | 2 |
| Robotics | 1 | 0 | 0 | 1 |
| Nanoscience | 0 | 1 | 0 | 1 |
| **Multiple** | **19 (11%)** | **4 (5%)** | **12 (4%)** | **35 (6%)** |
| **No specific field** | **89 (52%)** | **42 (56%)** | **93 (28%)** | **224 (38%)** |
| **[Cannot be attributed]** | **12 (7%)** | **0 (%)** | **9 (3%)** | **21 (4%)** |
| ***Total*** | ***170 (100%)*** | ***75 (100%)*** | ***338 (100%)*** | ***583 (100%)*** |

*Scientific field incidence (aggregated and disaggregated) by sub-corpus*

# Appendix 3: Original quotes in Italian

| Note number | Original Italian text |
|---|---|
| 14 | Nel concorso annullato di Oncologia clinica, due vincitori avevano un IF di 36 e 24 punti, contro i 359 del professor Robin Foà, bocciato. In quello di Medicina interna (ancora sub iudice), al candidato Pandolfi, in assoluto il migliore con 130 punti di IF, è stato preferito il suo collega Aliberti, che di punti ne aveva solo otto. |
| 15 | Il loro impact factor (il punteggio assegnato in base alle pubblicazioni scientifiche, molto importante per il giudizio della commissione) era di |
| | circa dieci volte inferiore di quello di Picano (in realtà anche qualcuno della commissione aveva un fattore minore di quello del candidato). |
| 20 | La ricerca, appena pubblicata sulla prestigiosa rivista Cerebral Cortex (Impact Factor 8.3), presenta il primo dato al mondo di una specifica variazione genetica correlata al deficit visivo associato alla dislessia. |
| 21 | Ora è arrivata la consacrazione del nuovo approccio, con la pubblicazione di un articolo sul "Cancer Journal for Clinicians", la rivista con il maggiore "impact factor" al mondo. |
| 23 | Se consideriamo l'*impact factor*, che misura il valore scientifico di una rivista con il numero di citazioni ricevute, quello di "Nature" è 40, mentre "Scientific Reports" supera di poco 4. |
| 24 | Giro il mondo e sono abituato ad essere valutato solo per le mie capacità scientifiche. Ho un H-Index di 60, Einstein aveva 100. Vuol dire che come ricercatore sono arrivato sull'Everest. |
| 29 | In pochi minuti è possibile creare la graduatoria di merito di tutto un dipartimento universitario togliendosi la soddisfazione di capire, sulla base di fattori perfettibili ma oggettivi, come si viene valutati dalla comunità di riferimento. Come minimo potrete includere il fattore H nel vostro curriculum o confrontarlo con quello dei commissari del vostro prossimo concorso. |
| 43 | Adesso sotto esame finiscono i professori e le loro ricerche: quello che scrivono, quello che pubblicano, gli studi degli ultimi sette anni. Sapremo come sta la ricerca in Italia e i voti che prenderanno i dipartimenti e le università nelle varie aree disciplinari. Una fotografia con il grandangolo puntato sulla produzione scientifica e umanistica. Si chiama "sistema di valutazione" e finirà con una "pagella" in base alla quale gli atenei migliori saranno premiati al momento della distribuzione delle risorse. |

# Supplementary materials

The database with the coded articles is available at XXXX [Database_articoli.xlsx]